\documentclass[final]{article}
\usepackage[latin1]{inputenc}
\usepackage[swedish,english]{babel}
\usepackage{graphicx}
\usepackage{amssymb}
\usepackage{array}
\usepackage{amsmath}
\usepackage{epstopdf}
\usepackage{epsfig}
\usepackage{subfigure}
\usepackage{pstricks}
\usepackage{fancyheadings}
\usepackage{color}

\newtheorem{Theorem}{Theorem}
\newtheorem{Corollary}{Corollary}

\newtheorem{Lemma}{Lemma}

\begin{document}
\title{High Order Finite Difference Schemes for the Elastic Wave Equation in Discontinuous Media}
\author{Kristoffer Virta$^{1,*}$, Kenneth Duru$^{2}$\\$^{1}$ Division of Scientific Computing,\\
Department of Information Technology, Uppsala.\\
$^2$ Department of Geophysics Stanford University, Stanford, CA.\\
$^{*}$Email: kristoffer.virta@it.uu.se}

\date{}
\maketitle

\pagenumbering{arabic}

\begin{abstract}
Finite difference schemes for the simulation of elastic waves in materials with jump discontinuities are presented. The key feature is the highly accurate treatment of interfaces where media discontinuities arise. The schemes are constructed using finite difference operators satisfying a summation - by - parts property together with a penalty technique to impose interface conditions at the material discontinuity. Two types of operators are used, termed fully compatible or compatible. Stability is proved for the first case by bounding the numerical solution by initial data in a suitably constructed semi - norm. Numerical experiments indicate that the schemes using compatible operators are also stable. However, the numerical studies suggests that fully compatible operators give identical or better convergence and accuracy properties. The numerical experiments are also constructed to illustrate the usefulness of the proposed method to simulations involving typical interface phenomena in elastic materials. 
\end{abstract}

%
\section{Introduction}
\label{sec:s1}
The elastic wave equation governs the propagation of seismic waves resulting from earthquakes and other seismic events. Other applications include waves in plates, beams and solid material structures. In the general setting the media can be described by piecewise smooth functions with jump discontinuities. This is especially true in seismological problems due to the layered structure of the earth. Large contrasts in media parameters may also be found in solid mechanics devices composed of different materials in welded contact. The presence of material discontinuities gives rise to reflection and refraction phenomena. For example earlier than expected arrivals of seismic waves, so called refraction arrivals, have been shown to exist in the presence of a material discontinuity \cite{cite:c17, cite:c2}. Similar to the Rayleigh surface wave \cite{cite:Rayleigh_1} an analogous Stoneley interface wave may under certain circumstances travel along media interfaces \cite{cite:Stoneley_1}. Also, as in the case of the traction free boundary a distinguishing characteristic of wave - interface interaction is that mode conversion occurs. That is, an incident wave type, either pressure or shear, is converted into two wave types, pressure and shear on reflection and refraction \cite{cite:Graff_1}, pp 377 - 380. In this paper, we consider the following problem: two elastic 2 dimensional half - planes are in contact at a line interface. We seek an approximation of the displacement field due to an initial disturbance as a function of time. We are particularly interested in the numerical simulation of the phenomena which arise from the influence of the line interface. 

To arrive at a solution to the problem we construct finite difference approximations of the elastic wave equation. The finite difference method has proven to be an efficient and easy to implement technique to approximate the elastic wave equation. However, the presence of jump discontinuities in media parameters makes the design of stable and accurate numerical methods challenging. It has been shown that smoothing or ignoring the material discontinuities in a numerical approximation reduces the formal order of accuracy to one \cite{cite:Brown_1}. That is, in order to preserve high order accuracy interfaces of jump discontinuities must be given special treatment. 

When using finite differences for simulation of wave propagation problems it has been known for a long time that high order methods (higher than 2) are superior to lower order methods \cite{cite:Kreiss_1}. It is also well known that stability together with accuracy guaranties convergence of the numerical scheme \cite{cite:Gustafsson_1}, pp 170.  Therefor a minimal requirement of the numerical scheme is high order accuracy and stability. 

To arrive at the approximate solution we use summation - by - parts (SBP) finite difference operators \cite{cite:c3,cite:c4} to approximate spatial derivatives. To maintain high accuracy in the presence of material discontinuities the required conditions at media interfaces are enforced with the simultaneous - approximation - term (SAT) methodology \cite{cite:Carpenter_1}. The elastic wave equation contains both second derivative terms such as $\partial/\partial x \left(b \partial u \partial x\right)$ and mixed derivative terms like $\partial/\partial x \left(b \partial u \partial y\right)$, where $b > 0$. To discretize these terms approximations of both first and second derivatives are required. Then certain compatibility conditions of the two approximations needs to be fullfilled. We use SBP approximations of first and second derivatives that are either compatible or fully compatible, the definitions will become clear in Section \ref{sec:s3}. This results in the case of fully compatible operators in a scheme which is proved to be stable in the sense that the corresponding approximative solution is bounded in a semi - norm by initial data. Stability in the case of compatible SBP operators is investigated by numerical experiments. The experiments indicate stability also in this case.

Discretizing the elastic wave equation has been done in several previous works. Benchmark problems which includes liquid  - solid interfaces was considered in \cite{cite:Stephen_1}. There material interfaces were not specified by any numerical boundary conditions. In \cite{cite:Virieux_1} the equations were rewritten to a first order system (1st order velocity - stress formulation) prior to discretizing. However, stability could only be shown for homogeneous materials. A recent and promising strategy includes the use of ghost points to enforce the correct conditions at the interface is described in \cite{cite:Petersson_1}. In that paper focus was put on the stable treatment of hanging nodes in a grid refinement and correct discretization of singular sources. Typical features of a discontinuous media was only considered in one experiment where a finite layer was put on top of a infinite half - space. In that experiment the results showed good agreement with a semi - analytical solution. The elastic wave equation with the traction free boundary condition was discretized using the SBP - SAT approach in \cite{DuruKreissMattsson2012}. The present study should be seen as a direct continuation of the work in \cite{DuruKreissMattsson2012}.

The general problem of seismic wave propagation includes non - planar boundaries. In the finite difference framework a coordinate transformation to curvilinear coordinates is therefor made before discretizing the equations. The general media might also be anisotropic. The theory in this paper is therefor presented in the general case of anisotropic media and in curvilinear coordinates. In order to focus on interface phenomena the numerical experiments are all carried out in isotropic media with piecewise constant media parameters and includes examples of mode conversion, refraction arrivals and the Stoneley interface wave. However, with the theory developed the generalization to curved geometries and anisotropic media is straightforward. 

Using an unstructured mesh greatly simplifies the handling of complicated geometries. In particular if hanging nodes are allowed. Methods with this approach includes discontinuous Galerkin \cite{cite:Kaser_1} and finite element \cite{cite:Cohen_1} discretizations of the elastic wave equation. Generating a high quality unstructured mesh can however be a non - trivial problem in itself and also requires extra bookkeeping and additional memory to keep track of the connectivity of the grid. It would therefor be interesting to study the performance of the method proposed in this paper on problems that includes curved geometries.                                 
          
The rest of the paper is outlined as follows. In Section \ref{sec:s2} the equations and interface conditions are introduced in the general setting. Section \ref{sec:s3} introduces the necessary definitions used to describe the discretization. Stability of the numerical scheme using fully compatible SBP operators is also proved here. In Section \ref{sec:s4} numerical experiments are presented. The purpose of the experiments are twofold. One, to verify stability and accuracy, two, to illustrate the usefulness of the proposed method on simulation of phenomena typical to wave motion in discontinuous elastic media. In particular, numerical examples of mode conversion, refraction arrivals and the Stoneley interface wave will be given. Section \ref{sec:s5} concludes and mentions future work.    
\section{Two elastic half - planes in contact}
\label{sec:s2}
Consider waves propagating in two elastic half - planes. The half - planes are in contact at $y = 0$. We denote the displacement field $U = (u_1,u_2)^T$ in the upper half-plane $\Omega = \left(-\infty, \infty\right) \times [0, \infty)$ and $U^\prime = (u^\prime_1, u^\prime_2)^T$ in the lower half-plane $\Omega' = \left(-\infty, \infty\right) \times (-\infty, 0]$. Let the positive quantities $\rho$, $c_{11}, c_{12} , c_{22}, c_{33}$ and $\rho^\prime$, $c_{11}^\prime, c_{12}^\prime , c_{22}^\prime, c_{33}^\prime$ be the density and elastic coefficients in $\Omega$ and $\Omega'$, respectively. The elastic coefficients may also include metric terms coming from a transformation to curvilinear coordinates. The elastic coefficients satisfy 
\begin{equation*}
	\label{eq:s2e5}
	c_{11}c_{22} - c_{12}^2 > 0, c'_{11}c'_{22} - c_{12}^{\prime 2} > 0.
\end{equation*}
The wave motion is governed by
\begin{equation}
	\label{eq:s2e1}
	\begin{array}{ll}
 		\rho U_{tt} =  \left( A U_x \right)_x  + \left(B U_y \right)_y   + \left(C U_y \right)_x  + \left(C^T U_x \right)_y, (x,y) \in \Omega,\\
	\rho^\prime U^\prime_{tt} =  \left(A' U^\prime_x\right)_x  + \left(B' U^\prime_y \right)_x + \left(C' U^\prime_y \right)_x  + \left(C'^T U^\prime_x \right)_y, (x,y) \in \Omega',
	\end{array}
	t > 0,
\end{equation}
where
\begin{equation*}
\label{eq:s2e2}
   A = \begin{bmatrix} 
	c_{11}&0\\
	0&c_{33}\\
	\end{bmatrix},
   B = \begin{bmatrix} 
	c_{33}&0\\
	0&c_{22}\\
	\end{bmatrix},
   C = \begin{bmatrix} 
	0&c_{12}\\
	c_{33}&0\\
	\end{bmatrix}.
\end{equation*}
The matrices $A^\prime, B^\prime$ and $C^\prime$ are defined analogously by the corresponding primed entries. In the following, we let unprimed quantities be defined in $\Omega$ and primed primed quantities be defined in $\Omega'$. When expressions are identical except for primed quantities we only write down the unprimed version. The solution $U$ is subject to initial data
\begin{equation}
	\label{eq:s2e3}
	\begin{array}{ll}
		U(x,y,0) = U_0, U_t (x,y,0) = U_1, (x,y) \in \Omega.\\
	\end{array}
\end{equation}
Here $U_0 \in L^2\left(\Omega\right), U_1 \in H^1_0\left(\Omega\right)$. From the solution of \eqref{eq:s2e1} we can compute normal and tangential stresses $\left(\tau_{22}, \tau_{12}\right)^T$ 
\begin{equation*}
	\begin{array}{ll}
		\left(\tau_{22}, \tau_{12}\right)^T = B U_y  + C^T U_x.  
	\end{array}
\end{equation*}
At the interface $y = 0$ continuity of normal stresses, tangential stresses and displacements are required,
\begin{equation}
	\label{eq:s2e4}
	\begin{array}{ll}
 		\tau_{22} - \tau_{22}' = 0,\\
		\tau_{12} - \tau_{12}' = 0,\\
	        U - U^\prime = 0,
  	\end{array}
	-\infty < x < \infty, y = 0.
  \end{equation}
Define the potential energy matrix
\begin{equation*}
	\label{eq:s2e7}
	P =\begin{bmatrix} 
		A&C\\
		C^T&B\\
	\end{bmatrix}.
\end{equation*}
Note that $P$ is symmetric positive semi - definite, see \cite{DuruKreissMattsson2012}. We define the elastic energies by
\begin{equation*}
	\label{eq:s2e8}
	\begin{array}{ll}
 	E\left(t\right) : = \int_{-\infty}^\infty \int_0^\infty \rho U_t^T  U_t dx dy + \int_{-\infty}^\infty \int_0^\infty \begin{bmatrix} 
	U_x\\
	U_y\\
\end{bmatrix}^T
P
 \begin{bmatrix} 
U_x\\
U_y\\
\end{bmatrix} dx dy, \\
	E^\prime\left(t\right) : = \int_{-\infty}^\infty \int_{-\infty}^0 \rho^{\prime} U^{\prime T}_t U^\prime_t dx dy + \int_{-\infty}^\infty \int_{-\infty}^0 \begin{bmatrix} 
	U'_x\\
	U^\prime_y\\
\end{bmatrix}^T
P^\prime
 \begin{bmatrix} 
 U^\prime_x\\
 U^\prime_y\\
\end{bmatrix} dx dy.
\end{array}
\end{equation*}
The total elastic energy is given by the sum of the elastic energies
\begin{align*}
E_T \left(t\right)= E\left(t\right) + E^\prime\left(t\right).
\end{align*}
It is straightforward to show that the elastic wave equation \eqref{eq:s2e1} with the  interface conditions \eqref{eq:s2e4} satisfy
\begin{equation*}
	\label{eq:s2e9}
 	E_T\left(t\right) = E_T\left(0\right), t \leq 0.
\end{equation*}
Hence, the total elastic energy is conserved. Thus, \eqref{eq:s2e1} together with \eqref{eq:s2e3} - \eqref{eq:s2e4} is a well posed problem.
\section{Spatial discretization}
\label{sec:s3}
%
%
In one spatial dimension the half - line $[0, \infty )$ is discretized by introducing an equidistant grid with grid size $h$
\begin{equation*}	
	\label{eq:s3e1}
	\begin{array}{ll}
	        x_j = j h,
	\end{array} j = 0,1,\dots.
\end{equation*} 
Let $u$ be a function defined on the half - line, the value $u(x_j)$ is denoted $u_{j}$. A one dimensional grid function corresponding to $u$ is then 
\begin{equation*}
	\mathbf{u} = \left(u_{0}, u_{1}, \dots\right).	
\end{equation*}
Let $\mathbf{u}$ and $\mathbf{v}$ be two grid functions. The scalar product of $\mathbf{u}$ and $\mathbf{v}$ is defined as usual by
\begin{equation*}
	<\mathbf{u}, \mathbf{v}> = \sum_{i=0}^{\infty} u_{i}v_{i}.
\end{equation*}
If $L$ is an operator acting on grid - functions its adjoint is denoted $L^*$.  
The first derivative $u_x$ is approximated by $D_1 \mathbf{u}$. Here $D_1$ is a $2p$-th order diagonal first derivative SBP operator,
\begin{equation}
	\label{eq:s3e2}
	\begin{array}{l}
		(D_1 \mathbf{u})_{j} = (u_x)_{j} + O(h^p), j = 0 \dots m_p,\\
		(D_1 \mathbf{u})_{j} = (u_x)_{j} + O(h^{2p}), j = m_p+1 \dots,\\
		D_1 = H^{-1} Q, \\
		<\mathbf{u}, \left(Q+Q^*\right)\mathbf{u}> = -n\mathbf{u}_0^2,\\
		<\mathbf{u}, H \mathbf{u}> = \sum_{j = 0}^\infty \mathbf{u}_{j}^2 h_j h, h_j > 0,
	\end{array}  
\end{equation}      
where the integer $m_p$ depends on $p$ and $n$ is the inward facing unit normal of the half - line. The SBP operator is termed diagonal due to the effect of multiplying each element of a grid function by a scalar when the operator $H$ is applied c.f., multiplying a diagonal matrix with a vector.  The variable coefficient second derivative $(b(x) u_x)_x,b(x) > 0$, is approximated by $D_2^{(b)}\mathbf{u}$, where $D_2^{(b)}$ is a $2p$-th order diagonal variable coefficient second derivative SBP operator \textit{compatible} with $D_1$,
\begin{equation}
	\label{eq:s3e3}
	\begin{array}{l}		
		(D_2^{(b)} \mathbf{u})_{j} = ((b u_x)_x)_{j} + O(h^p), j = 0 \dots n_p,\\
		(D_2^{(b)} \mathbf{u})_{j} = ((b u_x)_x)_{j} + O(h^{2p}), j = n_p+1 \dots,\\	
		D_2^{(b)} = H^{-1}(-M^{(b)} + \tilde{B}^{(b)}S), \\	
		M^{(b)} = D_1^* H B^{(b)} D_1 + R^{(b)} \geq 0,\\
		\left(B^{(b)} \mathbf{u}\right)_{j} = b_{j} \mathbf{u}_{j},\\	
		R^{(b)^*} = R^{(b)} \geq 0, \\
		(S\mathbf{u})_0 = (v_x)_0 + O(h^{p+1}),\\
		\left(\tilde{B}^{(b)} \mathbf{u}\right)_{j} = 
		\left\{ \begin{array}{ll}
				- n b_0 \mathbf{u}_0, j = 0,\\
				 0, j \neq 0,
			\end{array}
		\right.
	\end{array} 
\end{equation}
where the integer $n_p$ depends on $p$. The operator $H$ is the same in both definitions of $D_1$ and $D_2^{(b)}$. We also define a diagonal second derivative SBP operator \textit{fully compatible} with $D_1$ by requiring $S = D_1$ in \eqref{eq:s3e3}. Then one order of accuracy is lost at the point exactly on the boundary i.e., 
\begin{equation*}
	\begin{split}
		&(D_2^{(b)} \mathbf{u})_{0} = ((b u_x)_x)_{0} + O(h^{p-1}),\\
		&(S\mathbf{u})_0 = (v_x)_0 + O(h^{p}).
	\end{split}
\end{equation*} 
For details on SBP operators see e.g., \cite{cite:c3,cite:c4,cite:c8}. The grid, grid functions and operators corresponding to the half - line $(-\infty, 0]$ are defined analogously. 


We will also consider corresponding grid functions and operators on the line $\left(-\infty, \infty \right)$ discretized by
\begin{equation*}
 	x_j = j h, j = 0, \pm 1, \dots
\end{equation*} 
Now boundaries are absent and the corresponding SBP operators are defined by the interior schemes of \eqref{eq:s3e2} - \eqref{eq:s3e3}. To emphasize that it is the interior stencil that is used we use the notation $\bar{L}$ for operators acting on grid functions defined on the discretized line. Here $L$ is any operator acting on grid functions defined on the discretized half - line.  


\begin{figure}[htbp]
	\centering
	\includegraphics[scale = 0.8]{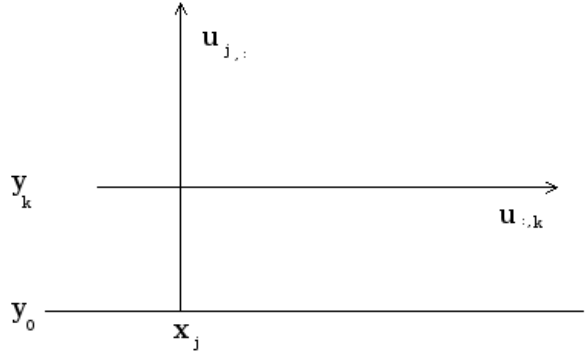}
	\caption{The function, grid function restricted to $x = h_j$ and $y = y_k$.}
	\label{fig:s3f1}
\end{figure}
In two spatial dimensions the upper half - plane is discretized by a two - dimensional equidistant grid with grid size $h$,
\begin{equation*}
	\label{eq:s3e4}
	\begin{array}{ll}
		x_j = j h, j = 0, \pm 1, \dots,\\
		y_k = k h, k = 0,1, \dots
	\end{array}
\end{equation*}
The value $u(x_j,y_k)$ of a function defined on the half-plane is denoted $u_{j,k}$ a two - dimensional grid function corresponding to $u$ is then $\mathbf{u}$ whose entry $\mathbf{u}_{j,k}$ is $u_{j,k}$. If $\mathbf{u}$ and $\mathbf{v}$ are two dimensional grid functions defined on the discretized half - plane the scalar product of $\mathbf{u}$ and $\mathbf{v}$ is computed as
\begin{equation*}
	\langle \mathbf{u},\mathbf{v} \rangle = \sum_{j=-\infty}^{\infty} \sum_{k=0}^\infty \mathbf{u}_{j,k} \mathbf{v}_{j,k}.
\end{equation*}
Fixing either the index $j$ or $k$ the one dimensional restrictions of $u$ or $\mathbf{u}$ to the points 
\begin{equation*}
	\label{eq:s3e5}
	\begin{array}{ll}
		\left(x_j,y_k\right), k = 0,1,\dots,\\
		\left(x_j,y_k\right), j = 0,\pm 1,\dots
	\end{array}
\end{equation*}
are denoted $u_{j,:}$, $u_{:,k}$ and $\mathbf{u}_{j,:}$, $\mathbf{u}_{:,k}$, respectively, see Figure \ref{fig:s3f1}. Let $L$ be an operator acting on one dimensional grid functions. We extend $L$ to two dimensional operators acting in the $x$ direction by repeated use of $\bar{L}$ on $\mathbf{u}_{:,k}$ and the $y$ direction by repeated use of $L$ in on $\mathbf{u}_{j,:}$. To distinguish whether an operator acting on two - dimensional grid functions is acting in the $x$ or $y$ direction a subscript $x$ or $y$ is used. That is,
\begin{equation*}
	\left(L_x \mathbf{u}\right)_{j,k} := \left(\bar{L} \mathbf{u}_{:,k}\right)_j, \left(L_y \mathbf{u}\right)_{j,k} := \left(L \mathbf{u}_{j,:}\right)_k.
\end{equation*}
The grid, grid functions and operators defined on a discretization of the lower half - plane are defined analogously.

We use the following operators to extract the boundary values of grid functions defined on either the upper or lower half plane and map it to a grid function on either the upper or lower half plane such that the only non - zeros values are located at the boundary,
\begin{equation*}
\label{eq:s3e6}
	 \begin{array}{ll}
		\left(E_{\pm2\pm}\mathbf{u}\right)_{0,k} = \mathbf{u}_{0,k},\\
		\left(E_{\pm2\pm}\mathbf{u}\right)_{j,k} = 0, j \neq 0.
	\end{array}
\end{equation*}
For example, the operator $E_{+2-}$ takes a grid function $\mathbf{u}$ defined on the upper discrete half - plane and maps it to a grid function defined on the lower half - plane which is zero everywhere except at the boundary, where it takes the corresponding values of $\mathbf{u}$. Note that
\begin{equation}
	\label{eq:s3e1112}
	E_{+2+} = E_{+2+}^*, E_{-2-} = E_{-2-}^*, E_{+2-} = E_{-2+}^*.
\end{equation}
Multiplication of grid functions is defined by element - wise multiplication. Multiplication from the left of a grid function and an operator is defined as
\begin{equation*}
	\label{eq:s3e7}
	\left(\mathbf{v}L\mathbf{u}\right)_{i,j} = \mathbf{v}_{i,j}\left(L\mathbf{u}\right)_{i,j}.
\end{equation*}   
\subsection{Spatial discretization of the equations}
We now discretize \eqref{eq:s2e1} together with \eqref{eq:s2e3} and \eqref{eq:s2e4} in the spatial coordinates. The result is a semi-discrete system in which time is the continuous variable. The discrete solutions are denoted $\mathbf{U} = \left(\mathbf{u}_1,\mathbf{u}_2\right)^T$ and $\mathbf{U}^\prime = \left(\mathbf{u}^\prime_1,\mathbf{u}^\prime_2\right)^T$. The right hand side of $\eqref{eq:s2e1}$ contains spatial derivatives of four types. These are discretized with the two - dimensional extensions of the SBP operators defined in \eqref{eq:s3e2} - \eqref{eq:s3e3} as
\begin{eqnarray*}
	\label{eq:s3e8}
		%
		%
		\begin{split}
		&\left(A U_x \right)_x \approx \begin{bmatrix} 
														 D_{2x}^{\left(c_{11}\right)} & 0 \\
														 0 & D_{2x}^{\left(c_{33}\right)}
												 \end{bmatrix}	
											         \mathbf{U} =: \mathbf{P}_{xx} \mathbf{U},\\
	        &\left(C U_y \right)_x \approx \begin{bmatrix} 
														 0 & D_{1x} \mathbf{c}_{12} D_{1y} \\
														 D_{1x} \mathbf{c}_{33} D_{1y} & 0
												 \end{bmatrix}	
											         \mathbf{U} =: \mathbf{P}_{yx} \mathbf{U},\\
	        &\left(B U_y \right)_y \approx \begin{bmatrix} 
														 D_{2y}^{\left(c_{33}\right)} & 0 \\
														 0 & D_{2y}^{\left(c_{22}\right)}
												 \end{bmatrix}	
											         \mathbf{U} =: \mathbf{P}_{yy} \mathbf{U},\\
	        &\left(C^T U_x \right)_y \approx \begin{bmatrix} 
														 0 & D_{1y} \mathbf{c}_{33} D_{1x} \\
														 D_{1y} \mathbf{c}_{12} D_{1x} & 0
												 \end{bmatrix}	
											         \mathbf{U} =: \mathbf{P}_{xy} \mathbf{U}.\\
	\end{split}
\end{eqnarray*}
Define
\begin{equation}
	\label{eq:s3e14}
	\mathbf{T} = 
		\begin{bmatrix}
			\mathbf{c}_{33} S_{1y} & \mathbf{c}_{33} D_{1x}\\
			\mathbf{c}_{12} D_{1x} & \mathbf{c}_{22} S_{1y}
		     \end{bmatrix}. 
\end{equation}
Normal and tangential stresses are then approximated by
\begin{equation}
	\label{eq:s3e9}
		     \begin{bmatrix}
			\tau_{22}\\
			\tau_{12}
		     \end{bmatrix} 
    			\approx 
		     \mathbf{T}
		     \mathbf{U} = \begin{bmatrix}
			\mathbf{t}_{22}\\
			\mathbf{t}_{12}
		     \end{bmatrix}.
\end{equation}
Introduce
\begin{equation}
	\label{eq:s3e10}
	\begin{array}{ll}
	\mathbf{Q} &= \mathbf{P}_{xx} + \mathbf{P}_{yy} + \mathbf{P}_{yx} + \mathbf{P}_{xy}.		 
	\end{array}
\end{equation}
The right hand side of \eqref{eq:s2e1} can then be discretized as
\begin{equation}
	\label{eq:s3e11}
	\begin{array}{ll}
	\left(A U_x \right)_x  + \left(B U_y \right)_y   + \left(C U_y \right)_x  + \left(C^T U_x \right)_y
\approx 
\mathbf{Q} \mathbf{U},\\
\left(A' U^\prime_x \right)_x  + \left(B' U^\prime_y \right)_y + \left(C' U^\prime_y \right)_x  + \left(C'^T U^\prime_x \right)_y 
\approx \mathbf{Q}' \mathbf{U}'.
\end{array}
\end{equation}  
With this notation the semi-discrete system may be written 
\begin{equation}
	\label{eq:s3e12}
	\begin{array}{ll}
		\mathbf{\rho} \mathbf{U}_{tt} = \mathbf{Q} \mathbf{U} + SAT,\\
		\mathbf{\rho}' \mathbf{U}'_{tt} = \mathbf{Q}' \mathbf{U}' + SAT'.
	\end{array}
\end{equation}
The semi-discrete solution is subject to initial data 
\begin{equation*}
	\label{eq:s3e131}
	\begin{array}{ll}
		\mathbf{U}(0) = \mathbf{U}_0, & \mathbf{U}_t(0) = \mathbf{U}_1,\\
		\mathbf{U}'(0) = \mathbf{U}'_0, & \mathbf{U}'_t(0) = \mathbf{U}'_1.
	\end{array}
\end{equation*}
In \eqref{eq:s3e12} the interface conditions \eqref{eq:s2e4} are imposed weakly via the penalty terms $SAT$ and $SAT'$. The penalty terms are proportional to the difference between discrete values of the displacements and stresses at the interfaces. 
Introduce 
\begin{equation}
	\label{eq:s3e13}
	\begin{array}{ll}
		SAT = SAT_{D} + SAT_{S},\\
		SAT' = SAT_{D}' + SAT_{S}'
	\end{array}
\end{equation}
and define
\begin{equation*}
	\label{eq:s3e141}
	\begin{array}{ll}
		I_{D_1} = E_{+2+}\mathbf{u}_1 - E_{-2+}\mathbf{u}_1', & I'_{D_1} = E_{-2-}\mathbf{u}'_1 - E_{+2-}\mathbf{u}_1,\\
		I_{D_2} = E_{+2+}\mathbf{u}_2 - E_{-2+}\mathbf{u}_2', & I'_{D_2} = E_{-2-}\mathbf{u}'_2 - E_{+2-}\mathbf{u}_2,\\
		I_{S_{22}} = E_{+2+}\mathbf{t}_{22} - E_{-2+}\mathbf{t}_{22}', & I_{S_{22}}' = E_{-2-}\mathbf{t}'_{22} - E_{+2-}\mathbf{t}_{22},\\
		I_{S_{12}} = E_{+2+}\mathbf{t}_{12} - E_{-2+}\mathbf{t}_{12}', & I_{S_{21}}' = E_{-2-}\mathbf{t}'_{12} - E_{+2-}\mathbf{t}_{12}.
	\end{array}
\end{equation*}
The penalty terms \eqref{eq:s3e13} can then be written as
\begin{equation}
	\label{eq:s3e15}
	\begin{split}
	&SAT_{D} =  \begin{bmatrix}
				 H_y^{-1} \left( \sigma_{11} I_{D_1} + \sigma_2 \left(\mathbf{c}_{33} S_{1y} \right)^* I_{D_1} + \sigma_2 H_x^{-1}\left(\mathbf{c}_{12} D_{1x} \right)^* H_x I_{D_2} \right)\\
				 H_y^{-1} \left( \sigma_{12} I_{D_2} + \sigma_2 H_x^{-1} \left( \mathbf{c}_{33} D_{1x} \right)^* H_x I_{D_1} + \sigma_2 \left(\mathbf{c}_{22} S_{1y}\right)^* I_{D_2} \right) 
			   \end{bmatrix}\\
	&SAT_{D}' =  \begin{bmatrix}
				 H_y^{-1} \left( \sigma_{11} I_{D_1} - \sigma_2 \left(\mathbf{c}_{33}' S_{1y} \right)^* I_{D_1}' - \sigma_2 H_x^{-1} \left(\mathbf{c}_{12}' D_{1x} \right)^* H_x I_{D_2}' \right)\\
				 H_y^{-1} \left( \sigma_{12} I_{D_2} - \sigma_2 H_x^{-1} \left( \mathbf{c}_{33}' D_{1x} \right)^* H_x I_{D_1}' - \sigma_2 \left(\mathbf{c}_{22}' S_{1y}\right)^* I_{D_2}' \right) 
			   \end{bmatrix}\\
	&SAT_{S} = 
	\begin{bmatrix}
		\sigma_3 H_y^{-1} I_{S_{22}}\\
		\sigma_3 H_y^{-1} I_{S_{12}}
	\end{bmatrix},\\
	&SAT_{S}' = 
	\begin{bmatrix}
		-\sigma_3 H_y^{-1} I_{S_{22}}'\\
		-\sigma_3 H_y^{-1} I_{S_{12}}'
	\end{bmatrix}.
	\end{split}
\end{equation} 
Here the terms $SAT_{D}$ and $SAT_{S}$ enforces the continuity of displacements and stresses, respectively. The penalty parameters $\sigma_{11}, \sigma_{12} , \sigma_2$ and $\sigma_3$ are determined below to yield a stable discretization of \eqref{eq:s2e1} together with \eqref{eq:s2e3} and \eqref{eq:s2e4}.
\subsection{Stability of the discretization}
Introduce the operators
\begin{equation*}
	\label{eq:s3e16}
	\begin{array}{lll}
	&\tilde{H} =
	\begin{bmatrix}
		H_x H_y & 0\\
		0 & H_x H_y
	\end{bmatrix},
	&\tilde{E}_{\pm2\pm} = 
	\begin{bmatrix}
		H_x E_{\pm2\pm} & 0\\
		0 & H_x E_{\pm2\pm}
	\end{bmatrix}.
	\end{array}
\end{equation*}
Using the properties \eqref{eq:s3e2} - \eqref{eq:s3e3} we can write \eqref{eq:s3e10} as
\begin{equation*}
	\begin{array}{ll}
		\mathbf{Q} = -\tilde{H}^{-1} \tilde{\mathbf{Q}} - \tilde{H}^{-1} \tilde{E}_{+2+} \mathbf{T}, & \mathbf{Q}' = -\tilde{H}^{-1} \tilde{\mathbf{Q}}' - \tilde{H}^{-1} \left(-\tilde{E}_{-2-}\right) \mathbf{T}'
	\end{array}
\end{equation*}
where
\begin{equation*}
	\label{eq:s3e161}
	\begin{split}
	&\tilde{\mathbf{Q}} = \tilde{\mathbf{P}}_{xx} + \tilde{\mathbf{P}}_{yy} + \tilde{\mathbf{P}}_{yx} + \tilde{\mathbf{P}}_{xy},\\
	&\tilde{\mathbf{P}}_{xx} = \begin{bmatrix}
		D_{1x}^* H_x H_y \mathbf{c}_{11} D_{1x} & 0 \\
		0 & D_{1x}^* H_x H_y \mathbf{c}_{33} D_{1x}
	\end{bmatrix} + \begin{bmatrix}
		H_y R_x^{\left(c_{11}\right)} & 0 \\
		0 &  H_y R_x^{\left(c_{33}\right)}
	\end{bmatrix},\\
	&\tilde{\mathbf{P}}_{yy} = \begin{bmatrix}
		D_{1y}^* H_x H_y \mathbf{c}_{33} D_{1y} & 0 \\
		0 & D_{1y}^* H_x H_y \mathbf{c}_{22} D_{1y}
	\end{bmatrix} + \begin{bmatrix}
		H_x R_y^{\left(c_{33}\right)} & 0 \\
		0 &  H_x R_y^{\left(c_{22}\right)}
	\end{bmatrix},\\
	&\tilde{\mathbf{P}}_{yx} = \begin{bmatrix} 
				0 & D_{1x}^* H_x H_y \mathbf{c}_{12} D_{1y} \\
		                D_{1x}^* H_x H_y \mathbf{c}_{33} D_{1y} & 0
			\end{bmatrix},\\
	&\tilde{\mathbf{P}}_{xy} = \begin{bmatrix} 
			  		0 & D_{1y}^* H_x H_y \mathbf{c}_{33} D_{1x} \\
	                       	        D_{1y}^* H_x H_y \mathbf{c}_{12} D_{1x} & 0
			 \end{bmatrix}
	.
	\end{split}
\end{equation*}
Note that
\begin{equation}
	\label{eq:s3e17}
\tilde{\mathbf{P}}_{xx} = \tilde{\mathbf{P}}_{xx}^*, \tilde{\mathbf{P}}_{yy} = \tilde{\mathbf{P}}_{yy}^*, \tilde{\mathbf{P}}_{yx} = \tilde{\mathbf{P}}_{xy}^*.  
\end{equation}
Define 
\begin{equation*}
	\label{eq:s3e18}
	   \begin{array}{ll} 
		\mathbf{B}_1 = \begin{bmatrix}
		        \tilde{E}_{+2+} & 0 \\
		        0 & \tilde{E}_{-2-}
		     \end{bmatrix},
		\mathbf{B}_2 = \begin{bmatrix}
		        0 & -\tilde{E}_{-2+} \\
		        -\tilde{E}_{+2-} & 0
		     \end{bmatrix},\\
		\mathbf{B}_3 = \begin{bmatrix}
		        \tilde{E}_{+2+} & 0 \\
		        0 & -\tilde{E}_{-2-}
		     \end{bmatrix},
		\mathbf{B}_4 = \begin{bmatrix}
		        0 & -\tilde{E}_{-2+} \\
		        \tilde{E}_{+2-} & 0
		     \end{bmatrix}.
	   \end{array}
\end{equation*}
the terms \eqref{eq:s3e15} can then be written as
\begin{equation*}
	\label{eq:s3e19}
	\begin{split}
		\begin{bmatrix}SAT_{D} \\ SAT_{D}'\end{bmatrix}  &= -\begin{bmatrix}\tilde{H} & 0 \\ 0 & \tilde{H} \end{bmatrix}^{-1} \left(-\begin{bmatrix}\sigma_{11} & 0\\0&\sigma_{12}\end{bmatrix} \left(\mathbf{B}_1 + \mathbf{B}_2\right)\begin{bmatrix}\mathbf{U}\\\mathbf{U}'\end{bmatrix}\right)\\
		&-\begin{bmatrix}\tilde{H} & 0 \\ 0 & \tilde{H} \end{bmatrix}^{-1} \left(-\sigma_2 \begin{bmatrix} \mathbf{T} & 0 \\ 0 & \mathbf{T}'\end{bmatrix}^* \left(\mathbf{B}_3 + \mathbf{B}_4\right)  \begin{bmatrix}\mathbf{U}\\\mathbf{U}'\end{bmatrix}\right),\\
		\begin{bmatrix}SAT_{S} \\ SAT_{S}'\end{bmatrix} &= -\begin{bmatrix}\tilde{H} & 0 \\ 0 & \tilde{H} \end{bmatrix}^{-1} \left(-\sigma_3\right) \left(\mathbf{B}_3 + \mathbf{B}_4\right) \begin{bmatrix} \mathbf{T} & 0 \\ 0 & \mathbf{T}'\end{bmatrix} \begin{bmatrix}\mathbf{U}\\\mathbf{U}'\end{bmatrix}.
	\end{split} 
\end{equation*}
With
\begin{equation}
	\label{eq:s3e20}
	\begin{split}
		\mathcal{M} &= \begin{bmatrix}\tilde{\mathbf{Q}} & 0 \\ 0 & \tilde{\mathbf{Q}'} \end{bmatrix} + \mathbf{B}_3 \begin{bmatrix} \mathbf{T} & 0 \\ 0 & \mathbf{T}'\end{bmatrix} -\begin{bmatrix}\sigma_{11} & 0\\0&\sigma_{12}\end{bmatrix} \left(\mathbf{B}_1 + \mathbf{B}_2\right)\\
	&- \sigma_2 \begin{bmatrix} \mathbf{T} & 0 \\ 0 & \mathbf{T}'\end{bmatrix}^* \left(\mathbf{B}_3 + \mathbf{B}_4\right) - \sigma_3 \left(\mathbf{B}_3 + \mathbf{B}_4\right) \begin{bmatrix} \mathbf{T} & 0 \\ 0 & \mathbf{T}'\end{bmatrix}
	\end{split}
\end{equation}
the system \eqref{eq:s3e12} can be written 
\begin{equation}
	\label{eq:s3e21}
	\begin{bmatrix} \rho \mathbf{U}\\ \rho' \mathbf{U}'\end{bmatrix}_{tt} = -\begin{bmatrix}\tilde{H} & 0 \\ 0 & \tilde{H} \end{bmatrix}^{-1} \mathcal{M} \begin{bmatrix}\mathbf{U}\\\mathbf{U}'\end{bmatrix}. 
\end{equation}
We now prove that the penalty parameters $\sigma_2$ and $\sigma_3$ can be chosen such that the operator $\mathcal{M}$ is self - adjoint. 
\begin{Lemma}[Self adjointness of the spatial operator]
	\label{lemma:s3l1}
	If $\sigma_2 = -\frac{1}{2}$, $\sigma_3 = \frac{1}{2}$, then $\mathcal{M}$ given by \eqref{eq:s3e21} is self - adjoint.
\end{Lemma}
\textbf{Proof:}\\
By \eqref{eq:s3e17} and \eqref{eq:s3e1112}
\begin{equation*}
		\tilde{\mathbf{Q}} = \tilde{\mathbf{Q}}^*, \mathbf{B}_1 = \mathbf{B}_1^*, \mathbf{B}_2 = \mathbf{B}_2^*, \mathbf{B}_3 = \mathbf{B}_3^*, \mathbf{B}_4 = -\mathbf{B}_4^*.
\end{equation*}
Write $\mathcal{M} = \mathcal{M}_1 + \mathcal{M}_2 + \mathcal{M}_3 + \mathcal{M}_4$ where
\begin{equation*}
	\mathcal{M}_1 = \begin{bmatrix}\tilde{\mathbf{Q}} & 0 \\ 0 & \tilde{\mathbf{Q}'} \end{bmatrix} = \begin{bmatrix}\tilde{\mathbf{Q}}^* & 0 \\ 0 & \tilde{\mathbf{Q}'}^* \end{bmatrix} = \mathcal{M}_1^*,
\end{equation*}
\begin{equation*}
	\mathcal{M}_2 = -\begin{bmatrix}\sigma_{11} & 0\\0&\sigma_{12}\end{bmatrix} \left(\mathbf{B}_1 + \mathbf{B}_2\right) = -\left(\mathbf{B}_1^* + \mathbf{B}_2^*\right)\begin{bmatrix}\sigma_{11} & 0\\0&\sigma_{12}\end{bmatrix}^* = \mathcal{M}_2^*, 
\end{equation*}
\begin{equation*}
	\begin{split}
		\mathcal{M}_3 &= \left(1-\sigma_3\right) \mathbf{B}_3 \begin{bmatrix} \mathbf{T} & 0 \\ 0 & \mathbf{T}'\end{bmatrix} - \sigma_2 \begin{bmatrix} \mathbf{T} & 0 \\ 0 & \mathbf{T}'\end{bmatrix}^* \mathbf{B}_3\\
		&= 1/2 \left( \mathbf{B}_3 \begin{bmatrix} \mathbf{T} & 0 \\ 0 & \mathbf{T}'\end{bmatrix} + \begin{bmatrix} \mathbf{T} & 0 \\ 0 & \mathbf{T}'\end{bmatrix}^* \mathbf{B}_3^* \right) = \mathcal{M}_3^*,
	\end{split} 
\end{equation*}
\begin{equation*}
	\begin{split}
	\mathcal{M}_4 &= \sigma_2 \begin{bmatrix} \mathbf{T} & 0 \\ 0 & \mathbf{T}'\end{bmatrix}^* \left(-\mathbf{B}_4\right) - \sigma_3 \mathbf{B}_4 \begin{bmatrix} \mathbf{T} & 0 \\ 0 & \mathbf{T}'\end{bmatrix}\\ &= -1/2 \left(\begin{bmatrix} \mathbf{T} & 0 \\ 0 & \mathbf{T}'\end{bmatrix}^* \mathbf{B}_4^* + \mathbf{B}_4 \begin{bmatrix} \mathbf{T} & 0 \\ 0 & \mathbf{T}'\end{bmatrix}\right) = \mathcal{M}_4^*.  
	\end{split}
\end{equation*}
Hence $\mathcal{M} = \mathcal{M}^*$ $\Box$
As a consequence of Lemma \ref{lemma:s3l1} we have the following corollary.
\begin{Corollary}[Conservation of energy]
\label{cor:s3c1}
All real-valued solutions $\left(\mathbf{U},\mathbf{U}'\right)$ to \eqref{eq:s3e21} satisfy 
\begin{equation}
	\label{eq:estimate}
	 \mathbf{E}(t) = C,
\end{equation}
where
\begin{equation}
	\label{eq:energy}
	\mathbf{E}(t) = \langle \begin{bmatrix}\mathbf{U} \\ \mathbf{U}'\end{bmatrix}_t, \begin{bmatrix}\rho \tilde{H} & 0\\0 & \rho' \tilde{H}\end{bmatrix} \begin{bmatrix}\mathbf{U} \\ \mathbf{U}'\end{bmatrix}_t \rangle  + \langle \begin{bmatrix}\mathbf{U} \\ \mathbf{U}'\end{bmatrix}, \mathcal{M} \begin{bmatrix}\mathbf{U} \\ \mathbf{U}'\end{bmatrix} \rangle
\end{equation}
and $C$ is a constant depending only on the initial data.
\end{Corollary}
\textbf{Proof:} 
Lemma \ref{lemma:s3l2} gives
\begin{equation*}
	\begin{split}
	&\frac{1}{2} \frac{d}{dt} \left(\langle \begin{bmatrix}\mathbf{U} \\ \mathbf{U}'\end{bmatrix}_t, \begin{bmatrix} \rho \tilde{H} & 0\\0 & \rho' \tilde{H}\end{bmatrix} \begin{bmatrix}\mathbf{U} \\ \mathbf{U}'\end{bmatrix}_t \rangle\right)\\
	&= -\frac{1}{2} \left(\langle \begin{bmatrix}\mathbf{U} \\ \mathbf{U}'\end{bmatrix}_t, \mathcal{M} \begin{bmatrix}\mathbf{U} \\ \mathbf{U}'\end{bmatrix} \rangle + \langle \mathcal{M}^*\begin{bmatrix}\mathbf{U} \\ \mathbf{U}'\end{bmatrix}, \begin{bmatrix}\mathbf{U} \\ \mathbf{U}'\end{bmatrix}_t \rangle \right)\\
	&= -\frac{1}{2} \frac{d}{dt} \langle \begin{bmatrix}\mathbf{U} \\ \mathbf{U}'\end{bmatrix}, \mathcal{M} \begin{bmatrix}\mathbf{U} \\ \mathbf{U}'\end{bmatrix} \rangle.  
	\end{split}
\end{equation*}
By integrating this expression in time we get \eqref{eq:estimate} $\Box$\\
For the quantity \eqref{eq:energy} to be an energy we need $\mathcal{M} \geq 0$. We will use the following calculations. 
\begin{equation}
	\label{eq:M1}
	\langle \begin{bmatrix}\mathbf{U} \\ \mathbf{U}'\end{bmatrix}, \mathcal{M}_1 \begin{bmatrix}\mathbf{U} \\ \mathbf{U}'\end{bmatrix} \rangle = \langle \mathbf{U},\tilde{\mathbf{Q}}\mathbf{U} \rangle + \langle \mathbf{U}',\tilde{\mathbf{Q}}'\mathbf{U}' \rangle.
\end{equation}
Here
\begin{equation}
	\label{eq:s3e111}
	\begin{split}
	\langle \mathbf{U},\tilde{\mathbf{Q}}\mathbf{U} \rangle &= \langle \begin{bmatrix}D_{1x} \mathbf{u}_1 \\ D_{1y} \mathbf{u}_2 \end{bmatrix},\tilde{H}\begin{bmatrix}\mathbf{c}_{11} & \mathbf{c}_{12} \\
	 \mathbf{c}_{12} & \mathbf{c}_{22}\end{bmatrix}\begin{bmatrix}D_{1x} \mathbf{u}_1 \\ D_{1y} \mathbf{u}_2 \end{bmatrix}\rangle \\
	&+ \langle \begin{bmatrix}D_{1y} \mathbf{u}_1 \\ D_{1x} \mathbf{u}_2 \end{bmatrix},\tilde{H}\begin{bmatrix}\mathbf{c}_{33} & \mathbf{c}_{33} \\ \mathbf{c}_{33} & \mathbf{c}_{33}\end{bmatrix}\begin{bmatrix}D_{1y} \mathbf{u}_1 \\ D_{1x} \mathbf{u}_2 \end{bmatrix}\rangle\\
	& + \langle \mathbf{u}_1,H_y R^{(c_{11})}_x \mathbf{u}_1 \rangle + \langle \mathbf{u}_2,H_y R^{(c_{33})}_x \mathbf{u}_2 \rangle \\
	&+ \langle \mathbf{u}_2,H_x R^{(c_{22})}_y \mathbf{u}_2 \rangle + \langle \mathbf{u}_1,H_x R^{(c_{33})}_y \mathbf{u}_1 \rangle \geq 0.
	\end{split}
	\end{equation}
The inequality holds since $c_{11} c_{22} - c_{12}^2 > 0$, $c_{33} > 0$ and \\$R_x^{(c_{11})}, R_x^{(c_{33})} R_x^{(c_{22})} R_x^{(c_{33})} \geq 0$. The first term of \eqref{eq:s3e111} can be computed as
\begin{equation*}
	\begin{split}
	&\langle \begin{bmatrix}D_{1x} \mathbf{u}_1 \\ D_{1y} \mathbf{u}_2 \end{bmatrix},\tilde{H}\begin{bmatrix}\mathbf{c}_{11} & \mathbf{c}_{12} \\
	 \mathbf{c}_{12} & \mathbf{c}_{22}\end{bmatrix}\begin{bmatrix}D_{1x} \mathbf{u}_1 \\ D_{1y} \mathbf{u}_2 \end{bmatrix}\rangle\\
	&=\sum_{i=-\infty}^{\infty} \sum_{j=0}^{\infty} \begin{bmatrix} \left(D_{1x}\mathbf{u}_1\right)_{i,j} \\ \left(D_{1y}\mathbf{u}_2\right)_{i,j} \end{bmatrix}^T \begin{bmatrix} \mathbf{c}_{11_{i,j}} & \mathbf{c}_{12_{i,j}} \\ \mathbf{c}_{12_{i,j}} & \mathbf{c}_{22_{i,j}}\end{bmatrix} \begin{bmatrix} \left(D_{1x}\mathbf{u}_1\right)_{i,j} \\ \left(D_{1y}\mathbf{u}_2\right)_{i,j} \end{bmatrix} h_i h_j h^2 \geq 0.
	\end{split}
\end{equation*}
Define 
\begin{equation*}
 \tilde{c}_i = \frac{1}{2} \left(\left(\mathbf{c}_{11_{i,0}} + \mathbf{c}_{22_{i,0}} \right) - \left(\left(\mathbf{c}_{11_{i,0}}-\mathbf{c}_{22_{i,0}}\right)^2 + 4\mathbf{c}_{12_{i,0}}^2\right)^{1/2}\right)
\end{equation*}
this is the smallest eigenvalue of the matrix
\begin{equation*}
	\begin{bmatrix} \mathbf{c}_{11_{i,0}} & \mathbf{c}_{12_{i,0}} \\ \mathbf{c}_{12_{i,0}} & \mathbf{c}_{22_{i,0}}\end{bmatrix}.
\end{equation*}
Since $c_{11} c_{22} - c_{12}^2 > 0$, $\tilde{c}_i > 0$ and
\begin{equation*}
	\begin{bmatrix} \mathbf{c}_{11_{i,0}} - \tilde{c}_i & \mathbf{c}_{12_{i,0}} \\ \mathbf{c}_{12_{i,0}} & \mathbf{c}_{22_{i,0}}- \tilde{c}_i\end{bmatrix} \geq 0.
\end{equation*}
We get
\begin{equation}
	\label{eq:M11}
	\begin{split}
	&\langle \begin{bmatrix}D_{1x} \mathbf{u}_1 \\ D_{1y} \mathbf{u}_2 \end{bmatrix},\tilde{H}\begin{bmatrix}\mathbf{c}_{11} & \mathbf{c}_{12} \\
	 \mathbf{c}_{12} & \mathbf{c}_{22}\end{bmatrix}\begin{bmatrix}D_{1x} \mathbf{u}_1 \\ D_{1y} \mathbf{u}_2 \end{bmatrix}\rangle \\
	&=\sum_{i=-\infty}^{\infty} \sum_{j=1}^{\infty} \begin{bmatrix} \left(D_{1x}\mathbf{u}_1\right)_{i,j} \\ \left(D_{1y}\mathbf{u}_2\right)_{i,j} \end{bmatrix}^T \begin{bmatrix} \mathbf{c}_{11_{i,j}} & \mathbf{c}_{12_{i,j}} \\ \mathbf{c}_{12_{i,j}} & \mathbf{c}_{22_{i,j}}\end{bmatrix} \begin{bmatrix} \left(D_{1x}\mathbf{u}_1\right)_{i,j} \\ \left(D_{1y}\mathbf{u}_2\right)_{i,j} \end{bmatrix} h_i h_j h^2\\
	&+ \sum_{i=-\infty}^{\infty} \begin{bmatrix}\left(D_{1x}\mathbf{u}_1\right)_{i,0} \\ \left(D_{1y}\mathbf{u}_2\right)_{i,0}\end{bmatrix}^T \begin{bmatrix} \mathbf{c}_{11_{i,0}} - \tilde{c}_i & \mathbf{c}_{12_{i,0}} \\ \mathbf{c}_{12_{i,0}} & \mathbf{c}_{22_{i,0}}- \tilde{c}_i\end{bmatrix} \begin{bmatrix} \left(D_{1x}\mathbf{u}_1\right)_{i,0} \\ \left(D_{1y}\mathbf{u}_2\right)_{i,0} \end{bmatrix} h_i h_0 h^2\\
	&+ \sum_{i=-\infty}^{\infty} \tilde{c}_i \left(D_{1x} \mathbf{u}_1\right)_{i,0}^2 h_i h_0 h^2 + \tilde{c}_i \left(D_{1y} \mathbf{u}_2\right)_{i,0}^2 h_i h_0 h^2 \geq 0.		
	\end{split}
\end{equation}
Similarly the second term of \eqref{eq:s3e111} can be computed as
\begin{equation}
	\label{eq:M12}
	\begin{split}
		&\langle \begin{bmatrix}D_{1y} \mathbf{u}_1 \\ D_{1x} \mathbf{u}_2 \end{bmatrix},\tilde{H}\begin{bmatrix}\mathbf{c}_{33} & \mathbf{c}_{33} \\ \mathbf{c}_{33} & \mathbf{c}_{33}\end{bmatrix}\begin{bmatrix}D_{1y} \mathbf{u}_1 \\ D_{1x} \mathbf{u}_2 \end{bmatrix}\rangle\\
		&=\sum_{i=-\infty}^{\infty} \sum_{j=1}^{\infty} \begin{bmatrix} \left(D_{1y}\mathbf{u}_1\right)_{i,j} \\ \left(D_{1x}\mathbf{u}_2\right)_{i,j} \end{bmatrix}^T \begin{bmatrix} \mathbf{c}_{33_{i,j}} & \mathbf{c}_{33_{i,j}} \\ \mathbf{c}_{33_{i,j}} & \mathbf{c}_{33_{i,j}}\end{bmatrix} \begin{bmatrix} \left(D_{1y}\mathbf{u}_1\right)_{i,j} \\ \left(D_{1x}\mathbf{u}_2\right)_{i,j} \end{bmatrix} h_i h_j h^2 \\
	&+ \sum_{i=-\infty}^{\infty} \mathbf{c}_{33_{i,0}} \left(\left(D_{1y} \mathbf{u}_1\right)_{i,0} +  \left(D_{1x} \mathbf{u}_2\right)_{i,0}\right)^2 h_i h_0 h^2 \geq 0.
	\end{split}
\end{equation}
The second term of \eqref{eq:M1} is computed analogously. We also compute
\begin{equation}
	\label{eq:M234}
	\begin{split}
	\langle \begin{bmatrix}\mathbf{U} \\ \mathbf{U}'\end{bmatrix}, \mathcal{M}_2 \begin{bmatrix}\mathbf{U} \\ \mathbf{U}'\end{bmatrix} \rangle & = \sum_{i=-\infty}^{\infty} 
\begin{bmatrix} \mathbf{u}_{1_{i,0}} & \mathbf{u}'_{1_{i,0}}\end{bmatrix} 
	\begin{bmatrix} 
		-\sigma_{11_{i,0}} & \sigma_{11_{i,0}}\\
		\sigma_{11_{i,0}} & -\sigma_{11_{i,0}}\\
	\end{bmatrix} 
	\begin{bmatrix} \mathbf{u}_{1_{i,0}} \\ \mathbf{u}'_{1_{i,0}}\end{bmatrix} h_i h_0 h\\
	&+\sum_{i=-\infty}^{\infty} 
	\begin{bmatrix} \mathbf{u}_{2_{i,0}} & \mathbf{u}'_{2_{i,0}}\end{bmatrix} 
	\begin{bmatrix} 
		-\sigma_{12_{i,0}} & \sigma_{12_{i,0}}\\
		\sigma_{12_{i,0}} & -\sigma_{12_{i,0}}\\
	\end{bmatrix} 
	\begin{bmatrix} \mathbf{u}_{2_{i,0}} \\ \mathbf{u}'_{2_{i,0}}\end{bmatrix} h_i h_0 h,\\	
	\langle \begin{bmatrix}\mathbf{U} \\ \mathbf{U}'\end{bmatrix}, \mathcal{M}_3 \begin{bmatrix}\mathbf{U} \\ \mathbf{U}'\end{bmatrix} \rangle &= \sum_{i = -\infty}^{\infty}
	\begin{bmatrix} \mathbf{u}_{1_{i,0}} & \mathbf{u}'_{1_{i,0}}\end{bmatrix} 
	\begin{bmatrix} 
		1 & 0 \\
		0 & -1
	\end{bmatrix} 
	\begin{bmatrix} \mathbf{t}_{22_{i,0}} \\ \mathbf{t}'_{22_{i,0}}\end{bmatrix} h_i h_0 h\\
	&+\sum_{i = -\infty}^{\infty}
	\begin{bmatrix} \mathbf{u}_{2_{i,0}} & \mathbf{u}'_{2_{i,0}}\end{bmatrix} 
	\begin{bmatrix} 
		1 & 0 \\
		0 & -1
	\end{bmatrix} 
	\begin{bmatrix} \mathbf{t}_{12_{i,0}} \\ \mathbf{t}'_{12_{i,0}}\end{bmatrix} h_i h_0 h\\
	\langle \begin{bmatrix}\mathbf{U} \\ \mathbf{U}'\end{bmatrix}, \mathcal{M}_4 \begin{bmatrix}\mathbf{U} \\ \mathbf{U}'\end{bmatrix} \rangle &= \sum_{i = -\infty}^{\infty}
	\begin{bmatrix} \mathbf{u}_{1_{i,0}} & \mathbf{u}'_{1_{i,0}}\end{bmatrix} 
	\begin{bmatrix} 
		0 & 1 \\
	       -1 & 0 \\
	\end{bmatrix} 
	\begin{bmatrix} \mathbf{t}_{22_{i,0}} \\ \mathbf{t}'_{22_{i,0}}\end{bmatrix} h_i h_0 h\\
	&+\sum_{i = -\infty}^{\infty} \begin{bmatrix} \mathbf{u}_{2_{i,0}} & \mathbf{u}'_{2_{i,0}}\end{bmatrix} 
	\begin{bmatrix} 
		0 & 1 \\
	       -1 & 0 \\
	\end{bmatrix} 
	\begin{bmatrix} \mathbf{t}_{12_{i,0}} \\ \mathbf{t}'_{12_{i,0}}\end{bmatrix} h_i h_0 h.
	\end{split}
\end{equation}
So far we have made no assumptions on the SBP operators used. We now prove that $\mathcal{M} \geq 0$ in the case of fully compatible SBP operators, that is with $S = D_1$ in the definition \eqref{eq:s3e3} of the diagonal second derivative SBP operators.
\begin{Lemma}[Ellipticity, case of fully compatible SBP operators]
	\label{lemma:s3l2}
	Using fully compatible SBP operators, all real - valued grid functions $\mathbf{U}$ and $\mathbf{U}'$
	\begin{equation*}
		\langle\begin{bmatrix}\mathbf{U} \\ \mathbf{U}'\end{bmatrix}, \mathcal{M} \begin{bmatrix}\mathbf{U} \\ \mathbf{U}'\end{bmatrix}\rangle \geq 0,
	\end{equation*} 	
if 	
	\begin{equation*}
	\begin{split}
		&\sigma_{22_{i,0}} < - \frac{\mathbf{c}_{33_{i,0}}}{4 h_0 h} - \frac{\mathbf{c}'_{33_{i,0}}}{4 h_0 h},\\
		&\sigma_{12_{i,0}} < - \frac{\mathbf{c}^2_{12_{i,0}}}{4 \tilde{c}_i h_0 h} - \frac{\mathbf{c}^2_{22_{i,0}}}{4 \tilde{c}_i h_0 h} - \frac{\mathbf{c}^{\prime 2}_{12_{i,0}}}{4 \tilde{c}_i h_0 h} - \frac{\mathbf{c}^{\prime 2}_{22_{i,0}}}{4 \tilde{c}'_i h_0 h}.
	\end{split}
\end{equation*}
\end{Lemma}
\textbf{Proof:}
We have
\begin{equation*}
	\langle\begin{bmatrix}\mathbf{U} \\ \mathbf{U}'\end{bmatrix}, \mathcal{M} \begin{bmatrix}\mathbf{U} \\ \mathbf{U}'\end{bmatrix}\rangle = \langle\begin{bmatrix}\mathbf{U} \\ \mathbf{U}'\end{bmatrix}, \mathcal{M}_1 \begin{bmatrix}\mathbf{U} \\ \mathbf{U}'\end{bmatrix}\rangle + \dots + \langle\begin{bmatrix}\mathbf{U} \\ \mathbf{U}'\end{bmatrix}, \mathcal{M}_4 \begin{bmatrix}\mathbf{U} \\ \mathbf{U}'\end{bmatrix}\rangle.
\end{equation*}
Here $\mathcal{M} = \mathcal{M}_1 + \mathcal{M}_2 + \mathcal{M}_3 + \mathcal{M}_4$ as in the proof of Lemma \ref{lemma:s3l2}.
Adding the terms of \eqref{eq:M1}, \eqref{eq:M234} and using \eqref{eq:M11} -\eqref{eq:M12} we get after some algebra.
\begin{align}
	\notag
		&\langle\begin{bmatrix}\mathbf{U} \\ \mathbf{U}'\end{bmatrix}, \mathcal{M} \begin{bmatrix}\mathbf{U} \\ \mathbf{U}'\end{bmatrix}\rangle 
	\notag
	=
		\sum_{i=-\infty}^{\infty} \sum_{j=1}^{\infty} \begin{bmatrix} \left(D_{1x}\mathbf{u}_1\right)_{i,j} \\ \left(D_{1y}\mathbf{u}_2\right)_{i,j} \end{bmatrix}^T \begin{bmatrix} \mathbf{c}_{11_{i,j}} & \mathbf{c}_{12_{i,j}} \\ \mathbf{c}_{33_{i,j}} & \mathbf{c}_{33_{i,j}}\end{bmatrix} \begin{bmatrix} \left(D_{1x}\mathbf{u}_1\right)_{i,j} \\ \left(D_{1y}\mathbf{u}_2\right)_{i,j} \end{bmatrix} h_i h_j h^2\\
	\notag
	&+ \sum_{i=-\infty}^{\infty} \sum_{j=1}^{\infty} \begin{bmatrix} \left(D_{1y}\mathbf{u}_1\right)_{i,j} \\ \left(D_{1x}\mathbf{u}_2\right)_{i,j} \end{bmatrix}^T \begin{bmatrix} \mathbf{c}_{33_{i,j}} & \mathbf{c}_{33_{i,j}} \\ \mathbf{c}_{33_{i,j}} & \mathbf{c}_{33_{i,j}}\end{bmatrix} \begin{bmatrix} \left(D_{1y}\mathbf{u}_1\right)_{i,j} \\ \left(D_{1x}\mathbf{u}_2\right)_{i,j} \end{bmatrix} h_i h_j h^2\\
	\notag
	&+\sum_{i=-\infty}^{\infty} \sum_{j=-\infty}^{1} \begin{bmatrix} \left(D_{1x}\mathbf{u}'_1\right)_{i,j} \\ \left(D_{1y}\mathbf{u}'_2\right)_{i,j} \end{bmatrix}^T \begin{bmatrix} \mathbf{c}'_{11_{i,j}} & \mathbf{c}'_{12_{i,j}} \\ \mathbf{c}'_{33_{i,j}} & \mathbf{c}'_{33_{i,j}}\end{bmatrix} \begin{bmatrix} \left(D_{1x}\mathbf{u}'_1\right)_{i,j} \\ \left(D_{1y}\mathbf{u}'_2\right)_{i,j} \end{bmatrix} h_i h_j h^2\\
	\notag
	&+ \sum_{i=-\infty}^{\infty} \sum_{j=-\infty}^{1} \begin{bmatrix} \left(D_{1y}\mathbf{u}'_1\right)_{i,j} \\ \left(D_{1x}\mathbf{u}'_2\right)_{i,j} \end{bmatrix}^T \begin{bmatrix} \mathbf{c}'_{33_{i,j}} & \mathbf{c}'_{33_{i,j}} \\ \mathbf{c}'_{33_{i,j}} & \mathbf{c}'_{33_{i,j}}\end{bmatrix} \begin{bmatrix} \left(D_{1y}\mathbf{u}'_1\right)_{i,j} \\ \left(D_{1x}\mathbf{u}'_2\right)_{i,j} \end{bmatrix} h_i h_j h^2\\
	\notag	
	&+ \sum_{i=-\infty}^{\infty} \begin{bmatrix}\left(D_{1x}\mathbf{u}_1\right)_{i,0} \\ \left(D_{1y}\mathbf{u}_2\right)_{i,0}\end{bmatrix}^T \begin{bmatrix} \mathbf{c}_{11_{i,0}} - \tilde{c}_i & \mathbf{c}_{12_{i,0}} \\ \mathbf{c}_{12_{i,0}} & \mathbf{c}_{22_{i,0}}- \tilde{c}_i\end{bmatrix} \begin{bmatrix} \left(D_{1x}\mathbf{u}_1\right)_{i,0} \\ \left(D_{1y}\mathbf{u}_2\right)_{i,0} \end{bmatrix} h_i h_0 h^2\\
	&+\sum_{i=-\infty}^{\infty} \begin{bmatrix} \mathbf{u}_{1_{i,0}} \\ \mathbf{u}_{1_{i,0}}' \\ \mathbf{t}_{22_{i,0}} \\ \mathbf{t}_{22_{i,0}}' \end{bmatrix}^T
	\label{eq:m1}
					\begin{bmatrix} -\sigma_{11_{i,0}} & \sigma_{11_{i,0}} & 1/2 & 1/2\\
					 \sigma_{11_{i,0}} & -\sigma_{11_{i,0}} & -1/2 & -1/2 \\
					  1/2 & -1/2 & \frac{h_0 h}{\mathbf{c}_{33_{i,0}}} & 0\\
					  1/2 & -1/2 & 0 & \frac{h_0 h}{\mathbf{c}'_{33_{i,0}}}
					\end{bmatrix} 
					\begin{bmatrix} \mathbf{u}_{1_{i,0}} \\ \mathbf{u}_{1_{i,0}}' \\ \mathbf{t}_{22_{i,0}} \\ \mathbf{t}_{22_{i,0}}' \end{bmatrix} h_i h \\
	\notag
	&+\sum_{i=-\infty}^{\infty} \begin{bmatrix} \mathbf{u}_{2_{i,0}} \\ \mathbf{u}_{2_{i,0}}' \\ \left(D_{1x} \mathbf{u}_1\right)_{i,0} \\ \left(D_{1y} \mathbf{u}_2\right)_{i,0} \\ \left(D_{1x} \mathbf{u}'_1\right)_{i,0} \\ \left(D_{1y} \mathbf{u}'_2\right)_{i,0}\end{bmatrix}^T\\
	\label{eq:m2}
	&\begin{bmatrix} -\sigma_{12_{i,0}} & \sigma_{12_{i,0}} & 1/2 & 1/2 & 1/2 & 1/2\\
					 \sigma_{12_{i,0}} & -\sigma_{12_{i,0}} & -1/2 & -1/2 & -1/2 & -1/2 \\
					  1/2 & -1/2 & \frac{\tilde{c}_1 h_0 h}{\mathbf{c}_{12_{i,0}}^2} & 0 & 0 & 0\\
					  1/2 & -1/2 & 0 & \frac{\tilde{c}_1 h_0 h}{\mathbf{c}_{22_{i,0}}^2} & 0 & 0\\
					  1/2 & -1/2 & 0 & 0 & \frac{\tilde{c}'_1 h_0 h}{\mathbf{c}_{12_{i,0}}^{\prime 2}} & 0\\
					  1/2 & -1/2 & 0 & 0 & 0 & \frac{\tilde{c}'_1 h_0 h}{\mathbf{c}_{22_{i,0}}^{\prime 2}}
					\end{bmatrix} 
					\begin{bmatrix} \mathbf{u}_{2_{i,0}} \\ \mathbf{u}_{2_{i,0}}' \\ \left(D_{1x} \mathbf{u}_1\right)_{i,0} \\ \left(D_{1y} \mathbf{u}_2\right)_{i,0} \\ \left(D_{1x} \mathbf{u}'_1\right)_{i,0} \\ \left(D_{1y} \mathbf{u}'_2\right)_{i,0}\end{bmatrix} h_i h\\
	\notag
	&+ \langle \mathbf{u}_1,R^{(c_{11})}_x \mathbf{u}_1 \rangle + \langle \mathbf{u}_2,R^{(c_{33})}_x \mathbf{u}_2 \rangle + \langle \mathbf{u}_2,R^{(c_{22})}_y \mathbf{u}_2 \rangle + \langle \mathbf{u}_1,R^{(c_{33})}_y \mathbf{u}_1 \rangle\\
	\notag
	&+ \langle \mathbf{u}_1,R^{(c_{11})}_x \mathbf{u}_1 \rangle + \langle \mathbf{u}_2,R^{(c_{33})}_x \mathbf{u}_2 \rangle + \langle \mathbf{u}_2,R^{(c_{22})}_y \mathbf{u}_2 \rangle + \langle \mathbf{u}_1,R^{(c_{33})}_y \mathbf{u}_1 \rangle.
\end{align}
The first five and the last eight terms of this sum are non - negative. The matrices of \eqref{eq:m1} and \eqref{eq:m2} also appeared in \cite{Virta_1}. There it was shown that if
\begin{equation*}
	\begin{split}
		&\sigma_{22_{i,0}} < - \frac{\mathbf{c}_{33_{i,0}}}{4 h_0 h} - \frac{\mathbf{c}'_{33_{i,0}}}{4 h_0 h},\\
		&\sigma_{12_{i,0}} < - \frac{\mathbf{c}_{12_{i,0}}}{4 \tilde{c}_i h_0 h} - \frac{\mathbf{c}_{22_{i,0}}}{4 \tilde{c}_i h_0 h} - \frac{\mathbf{c}'_{12_{i,0}}}{4 \tilde{c}'_i h_0 h} - \frac{\mathbf{c}'_{22_{i,0}}}{4 \tilde{c}'_i h_0 h}
	\end{split},
\end{equation*}
then the matrices are positive semi - definite.
Hence with this choice of parameters $\mathcal{M} \geq 0$ $\Box$\\
%
%
%
Lemma \ref{lemma:s3l1}, Corollary \ref{cor:s3c1} and Lemma \ref{lemma:s3l2} proves the main result of this paper,
\begin{Theorem}
	\label{thm:s3t1}
	The system \eqref{eq:s3e21} obtained with fully compatible SBP operators is stable in the sense that its solution is bounded by initial data in the semi - norm \eqref{eq:energy} if the penalty parameters $\sigma_{11}, \sigma_{12}, \sigma_{2}$ and $\sigma_3$ are chosen as in Lemma \ref{lemma:s3l1} and \ref{lemma:s3l2}.
\end{Theorem}
The proof that $\mathcal{M} \geq 0$ depends on $S = D_1$ in the approximation of the stresses by \eqref{eq:s3e14} and \eqref{eq:s3e9}. For the case of compatible SBP operators $S \neq D_1$ and the proof does not hold. Intuitively it would seem to be preferable to use compatible SBP operators, if such a discretization is stable. This is due to the fact that fully compatible SBP operators loses one order of accuracy in exactly one point at the boundary. In the next section we investigate both discretizations where it seems, although we can not prove it, that a scheme using fully compatible SBP operators is also stable. However, the numerical experiments also show, rather counterintuitive, that the schemes using fully compatible operators have identical or better accuracy and convergence properties. 
\section{Numerical experiments}
\label{sec:s4}
In this section we present experiments with the numerical method described above. In particular, we will use schemes that have been constructed using 4th and 6th order diagonal fully compatible and compatible SBP operators. We choose the material parameters to represent two anisotropic elastic medium. In particular, if $\lambda$ and $\mu$ are the first and second Lam\'e parameters, respectively. Then 
\begin{equation*}
	c_{11} = c_{22} = \lambda + 2\mu, c_{12} = \lambda, c_{33} = \mu. 
\end{equation*}
To integrate numerically in time we use the 4th order scheme described in \cite{cite:Gilbert_1} which was designed for systems on the form \eqref{eq:s3e21}. We will consider three experiments that serve as verification of stability and accuracy of the proposed method, as well as its applicability to interface phenomena in layered elastic materials.
\subsection{Mode conversion at a line interface}
\label{exp:e1}
\begin{figure}[htbp]
	\centering
	\includegraphics[scale = 0.50]{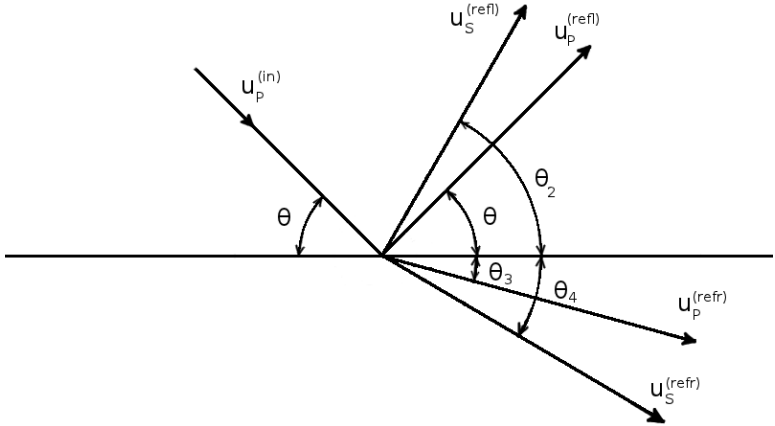}	
	\caption{Reflection and refraction of a plane pressure wave impinging on the interface between two elastic half-spaces.}
	\label{fig:s4f2}
\end{figure}

\begin{figure}[htbp]
	\centering
	\subfigure[4th order, compatible]{
		\includegraphics[scale = 0.35]{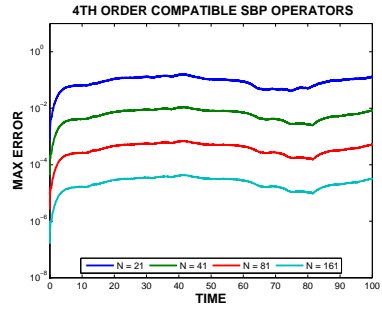}
		\label{fig:s4f1_b}	
	}
	\subfigure[6th order, compatible]{
		\includegraphics[scale = 0.35]{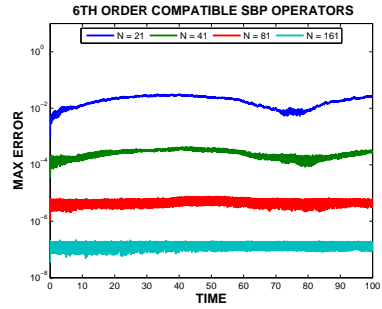}
		\label{fig:s4f1_c}	
	}
	\subfigure[4th order, fully compatible]{
		\includegraphics[scale = 0.35]{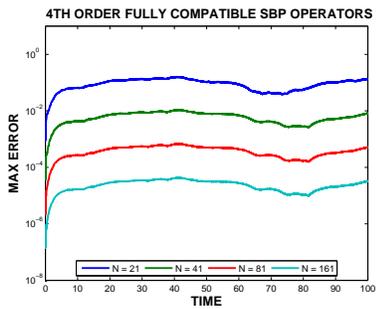}
		\label{fig:s4f1_e}	
	}
	\subfigure[6th order, fully compatible]{
		\includegraphics[scale = 0.35]{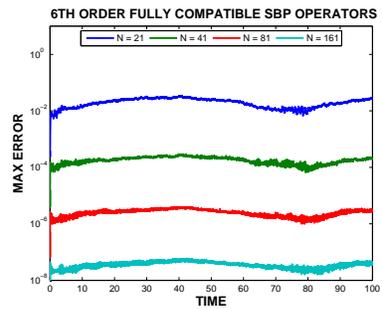}
		\label{fig:s4f1_f}	
	}
	\label{fig:s4f1}
	\caption{Max error as function of time for a plane wave impinging on an interface where discontinuities in the Lam\'e parameters occur. Results from methods using fourth and sixth order SBP operators are shown from left to right. The number of grid points increases from top to bottom.}
\end{figure}

In a first experiment we consider a compressional plane wave of unit amplitude propagating with angle $\theta$ and temporal frequency $\frac{\omega}{2\pi}$ in the negative $y$-direction. The displacement field is given by
\begin{equation*}
	\begin{split}
		&U_P^{(\mathrm{in})} = \begin{pmatrix} \xi \\ -\eta \end{pmatrix} e^{i(\gamma \xi x - \gamma \eta y - \omega t)},\\
		&\theta \in (0,\frac{\pi}{2}),\xi = \sin(\theta), \eta = \cos(\theta),\gamma = \frac{\omega}{\sqrt{\lambda+2\mu}}, 
	\end{split}
	x \in (-\infty, \infty), y \in [0, \infty).
\end{equation*}
We assume an interface between two different materials at $y = 0$. The Lame parameters and densities are $\lambda, \mu, \rho$ and $\lambda^\prime, \mu^\prime, \rho'$ in the half-planes $y>0$ and $y<0$, respectively.  
When this wave encounters the interface between the two different materials it will be split into reflected, compressional and shear waves,
\begin{eqnarray*}
	\begin{split}
		&U = U^{\mathrm{(in)}} + U^{\mathrm{(refl)}}_P + U^{\mathrm{(refl)}}_S,\\
		&U^{\mathrm{(refl)}}_P = A_{\mathrm{refl}} \begin{pmatrix} \xi \\ \eta \end{pmatrix} e^{i(\gamma \xi x + \gamma \eta y - \omega t)},\\
		&U^{\mathrm{(refl)}}_S = B_{\mathrm{refl}} \begin{pmatrix} \gamma_2 \eta_2 \\ \gamma \xi \end{pmatrix} e^{i(\gamma \xi x + \gamma_2 \eta_2 y - \omega t)},\\
		&\eta_2 = \cos(\theta_2), \sin(\theta_2) = \frac{\gamma}{\gamma_2} \xi, \gamma_2 = \frac{\omega}{\sqrt{\mu}}, 		
	\end{split}
	x \in (-\infty, \infty), y \in (\infty,0]
\end{eqnarray*}
and refracted compressional and shear waves,
\begin{equation*}
	\begin{split}
		&U' = U^{\mathrm{(refr)}}_P + U^{\mathrm{(refr)}}_S,\\
		&U^{\mathrm{(refr)}}_P = A_{\mathrm{refr}} \begin{pmatrix} \gamma_1 \xi_1 \\ \gamma_3 \eta_3 \end{pmatrix} e^{i(\gamma_1 \xi_1 x - \gamma_3 \eta_3 y - \omega t)},\\
		&\eta_3 = \cos(\theta_3), \sin(\theta_3) = \frac{\gamma_1}{\gamma_3} \xi_1,\gamma_3 = \frac{\omega}{\sqrt{\lambda^\prime + 2\mu^\prime}},\\
		&U^{\mathrm{(refr)}}_S = B_{\mathrm{refr}} \begin{pmatrix} -\gamma_4 \eta_4 \\ \gamma_1 \xi_1 \end{pmatrix} e^{i(\gamma_1 \xi_1 x - \gamma_4 \eta_4 y - \omega t)},\\
		&\eta_4 = \cos(\theta_4), \sin(\theta_4) = \frac{\gamma_1}{\gamma_4} \xi_1,\gamma_4 = \frac{\omega}{\sqrt{\mu^\prime}}, 
	\end{split}
	x \in (-\infty, \infty), y \in (-\infty,0],
\end{equation*}
see Figure \ref{fig:s4f2}. The constants $A_{\mathrm{refl}}, B_{\mathrm{refl}}, A_{\mathrm{refr}}, B_{\mathrm{refr}}$ are obtained by inserting $U$ and $U'$ into the conditions \eqref{eq:s2e4} and solving the resulting linear system. For a more detailed discussion see \cite{cite:Graff_1}, pp 377 - 380. We choose $\theta = \frac{\pi}{4}$, $\omega = 2\pi$, $\lambda = \mu = 1$, $\lambda^\prime = 1/2, \mu^\prime = 2$ and $\rho = \rho^\prime = 1$. We use this exact solution to verify stability and accuracy of the numerical schemes described in this work. The computational domain is taken to be $(x,y) \in [-\frac{2\pi}{\gamma \xi}, \frac{2\pi}{\gamma \xi}]\times [-\frac{2\pi}{ \gamma_2\eta_2},\frac{2\pi}{ \gamma_3\eta_3}]$. Initial data for the numerical scheme is taken as the real part of the analytic solution at time $t=0$ and exact data is imposed at the outer boundaries. The solution is computed for 100 periods until $t = 100$ and the discrete max - error is measured. We use $2 N \times N$ grid points. Figures \ref{fig:s4f1_b} - \ref{fig:s4f1_c} display the max error in the numerical solution as a function of time for $t \leq 100$ obtained with schemes using compatible SBP operators. Figures \ref{fig:s4f1_e} - \ref{fig:s4f1_f} display the corresponding measurements obtained with fully compatible SBP operators. These figures illustrate the stability and accuracy of the schemes. Numerical values of the rate of convergence measured at time $t = 100$ are shown in Table \ref{tab:s4t1}. Stability was only proved for the schemes using fully compatible SBP operators. The computations done with compatible SBP operators indicates that these schemes are also stable. Note that the computations for this example suggest that the 6th order fully compatible SBP operators yield schemes that have slightly better accuracy and convergence properties than the corresponding 6th order compatible versions even though the formal order of accuracy is one order higher at exactly one point for the compatible SBP operators.  
\begin{table}[htbp]
	\begin{center}
            	\begin{tabular}{|l|l|l|l|l|l|l|l|l|l|}
			\hline
			\multicolumn{3}{|c|}{Compatible SBP operators} & \multicolumn{2}{c|}{Fully compatible SBP operators}\\
			\hline
            		$N$ &  $p_4$ & $p_6$ &  $p_4$ & $p_6$  \\
			\hline
            		21 & -  & - & -  & - \\
			\hline
            		41 & 3.98 & 6.53  & 4.07 & 7.06\\
			\hline
			81 & 4.01 & 5.57  & 4.00 & 5.98\\
			\hline
			161 & 4.00 & 5.52  & 3.99 & 6.37\\
			\hline
            	\end{tabular}
		\caption{Rate of convergence of the solution in max norm. $p_4$ and $p_6$ gives the measured convergence using 4th and 6th order operators, respectively.}
            	\label{tab:s4t1}	
            \end{center}      
\end{table}
\subsection{Refraction arrivals}
\begin{figure}
	\centering
	\includegraphics[scale=0.12]{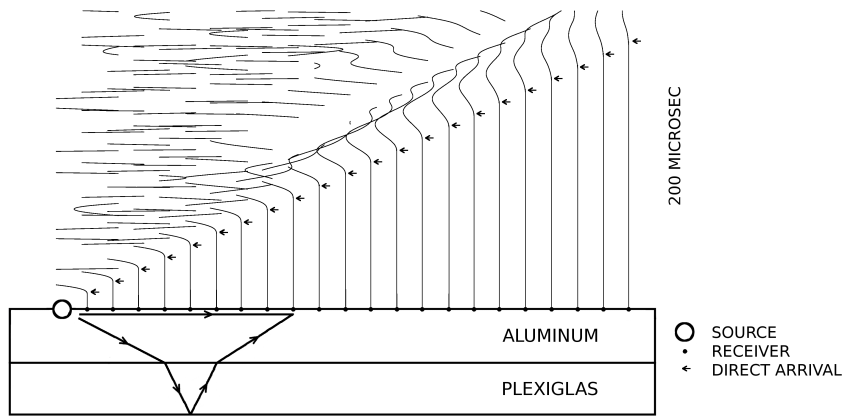}	
	\caption{The source is placed at the upper boundary of the plate of Plexiglas. A refraction arrival becomes apparent 12 inches from the source. The arrows from the source displays the path of the direct wave and the refracted wave, respectively.}
	\label{fig:s4f3}
\end{figure}

\begin{figure}
	\centering
	\includegraphics[scale=0.115]{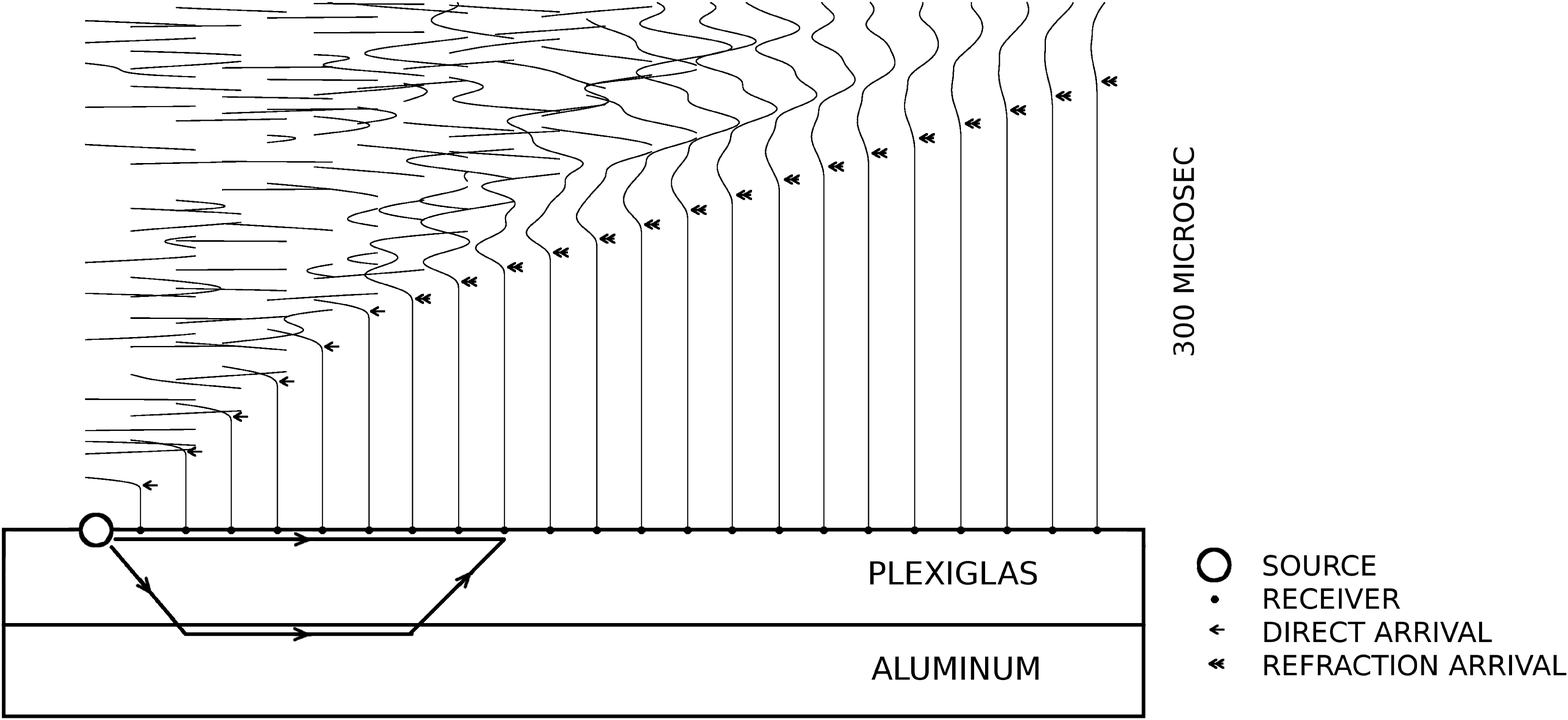}
	\caption{The source is placed at the upper boundary of the plate of aluminum. The direct wave is now the first arrival. The arrows from the source displays the path of the direct wave and the refracted wave, respectively.}
	\label{fig:s4f4}
\end{figure}

\begin{figure}
	\centering
	\includegraphics[scale=0.8]{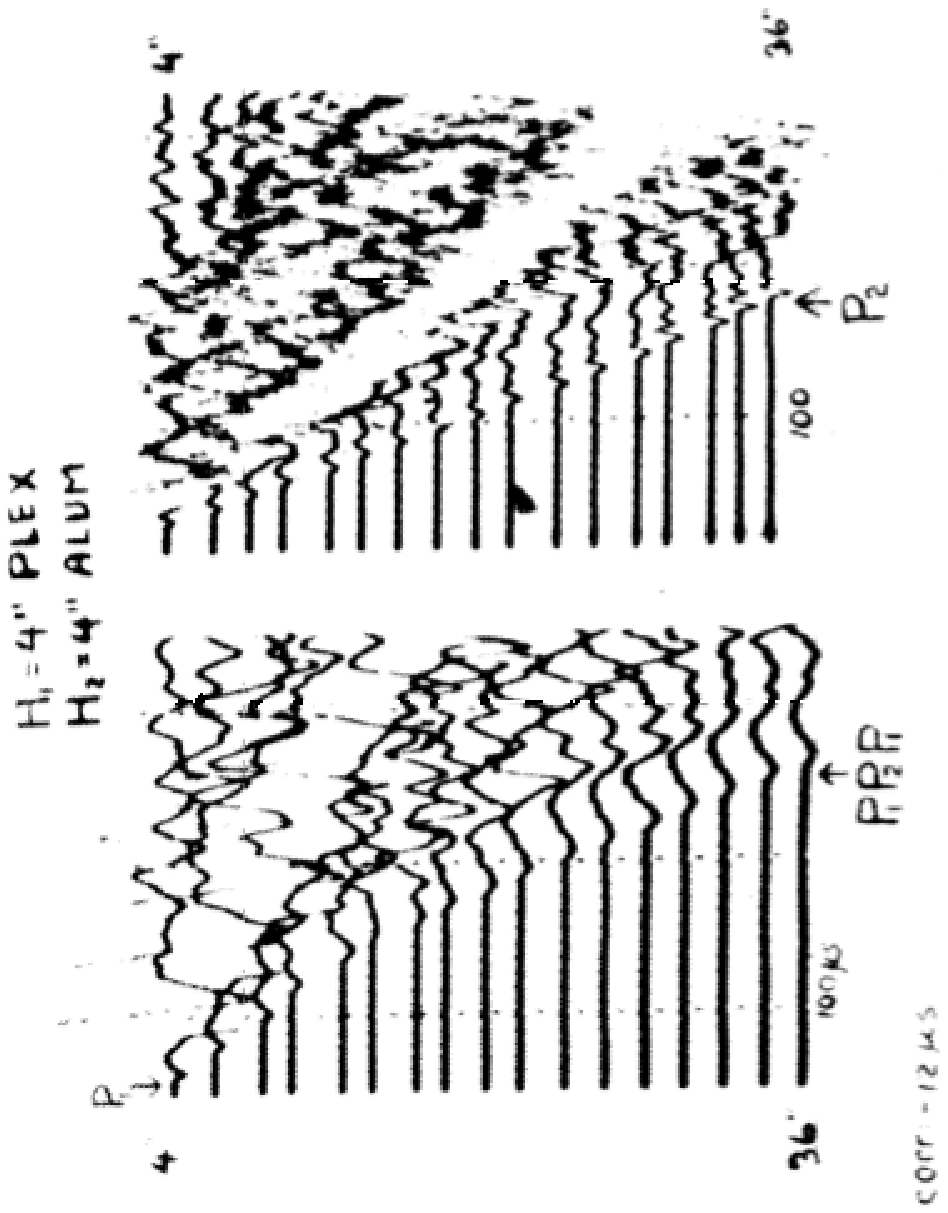}
	\caption{Results from the laboratory experiment presented in \cite{cite:c17}. Top figure: the source is placed at the upper boundary of the plate of aluminum. Bottom figure: the source is placed at the boundary of the plate of Plexiglas.}
	\label{fig:s4f5}
\end{figure}

Elastic wave motions can be significantly affected by a layered structure of the underlying media. We illustrate this by solving a version of Lamb\'s problem \cite{cite:Lamb_1} on a domain consisting of two materials in welded contact. The domain is composed of a plate of Plexiglas welded together with a plate of aluminum, both plates measure $50$ by $4$ inches. The plate of Plexiglas has pressure - wave velocity $C^{Pl}_p = 2630 m / s$, shear - wave velocity $C^{Pl}_s = 1195 m / s$ and density $\rho^{Pl} = 1190 kg / m^3$, the plate of aluminum has pressure -wave velocity $C^{Al}_p = 6460 m/s$, shear - wave velocity $C^{Al}_s = 3100 m/s$ and density $\rho^{Al} = 2700 kg / m^3$. Initially, the displacements and velocities are zero and the problem is forced by adding a source term 
\begin{equation*}
	f = 
	\left\{
	\begin{array}{ll}
		10^4 \left(\sin(2 \pi 250 t) - 1/2 \sin(2 \pi 500 t)\right)\delta(x-4), t \in \left(0, 1/250\right) \\
		0, \mathrm{else}
	\end{array} \right.
\end{equation*}
to the normal stresses at the upper horizontal traction free boundary of either the plate of Plexiglas or the plate of aluminum. The source term is located $4$ inches from the left corner of the upper plate, see Figure \ref{fig:s4f3}.  On the vertical boundaries a traction free boundary condition is imposed. The response is then recorded as a function of time by receivers spaced $2$ inches apart, see Figures \ref{fig:s4f3} and \ref{fig:s4f4}. A generalization of this problem in which two semi - infinite half - planes was considered is discussed in \cite{cite:c2}. There it is shown that refraction arrivals are generated if a source is placed in the half - space of lower pressure and shear - wave speeds as the wave front impinges on the interface to the faster media. Refraction arrivals was also measured experimentally in \cite{cite:c17} where a spark served as a point source on the boundary of a plate of Plexiglas welded together with a plate of aluminum or vice versa, as described above. Here refraction arrivals was observed only in the case of a source on the boundary of the slower plate of Plexiglas, in accordance with the theory in \cite{cite:c2}. The same behavior is observed in the present numerical experiment. In the case of the source located on the boundary of the plate of Plexiglas an earlier refraction arrival is measured at the receivers located $12$ inches and further away from the source. In the other case the direct wave is always the first arrival, see Figures \ref{fig:s4f3} and \ref{fig:s4f4}. For comparison the results of the corresponding laboratory experiment presented in \cite{cite:c17} are displayed in Figure \ref{fig:s4f5}. Features of the recordings are similar, in both experiments the refraction arrival first becomes apparent at the recorder spaced 12 inches away from the source.               
\subsection{The Stoneley interface wave}
A Stoneley interface wave satisfies the elastic wave equation in two - homogeneous half - planes welded in contact. Define
\begin{equation*}
	\label{eq:s4e3e0}
	\alpha = \sqrt{\frac{\lambda + 2 \mu}{\rho}}, \beta = \sqrt{\frac{\mu}{\rho}}.
\end{equation*}
To simplify we assume that the Stoneley wave is $2\pi$ - periodic in the $x$ - direction. The component of the Stoneley wave in the half  - plane $y \geq 0$ can then be written
\begin{equation}
	\label{eq:s4e3e1}
	\begin{split}
		U &= A e^{-y\sqrt{1-c_S^2 / \alpha^2}} \begin{pmatrix} \cos(x-c_S t) \\  \sqrt{1-c_S^2 / \alpha^2}\sin(x-c_S t) \end{pmatrix} \\
		  &+ B e^{-y\sqrt{1-c_S^2 / \beta^2}} \begin{pmatrix} -\sqrt{1-c_S^2 / \beta^2} \cos(x-c_S t) \\ -\sin(x-c_S t) \end{pmatrix},	
	\end{split}
	 y \geq 0.
\end{equation}
The component of the Stoneley wave in the half - plane $y \leq 0$ is written
\begin{equation}
	\label{eq:s4e3e2}
	\begin{split}
		U' &= C e^{y\sqrt{1-c_S^2 / \alpha^{\prime 2}}} \begin{pmatrix} \cos(x-c_S t) \\  -\sqrt{1-c_S^2 / \alpha^{\prime 2}}\sin(x-c_S t) \end{pmatrix} \\
		  &+ D e^{y\sqrt{1-c_S^2/ \beta^{\prime 2}}} \begin{pmatrix} \sqrt{1-c_S^2 / \beta^{\prime 2}} \cos(x-c_S t) \\ -\sin(x-c_S t) \end{pmatrix},	
	\end{split}
	 y \leq 0.
\end{equation}

As in Experiment \ref{exp:e1} the constants $A,B,C$ and $D$ are determined through the solution to the linear system arising form inserting \eqref{eq:s4e3e1} and \eqref{eq:s4e3e2} into the interface conditions \eqref{eq:s2e4}. This system now depends on the Stoneley phase velocity $c_S$ and a solution exists iff $c_S$ satisfies the dispersion relation
%


\begin{equation}
	\label{eq:s4e3e3}
	\det \begin{vmatrix} 1 & -\sqrt{1-\frac{c_S^2}{\beta^2}} & -1 & -\sqrt{1-\frac{c_S^2}{\beta^{\prime 2}}}\\
			     \sqrt{1-\frac{c_S^2}{\alpha^2}} & -1 & \sqrt{1-\frac{c_S^2}{\alpha^{\prime 2}}} & 1 \\
 -\rho c_S^2-2 \rho \beta^2  & 2 \rho \beta^2 \sqrt{1-\frac{c_S^2}{\beta^2}} & \rho'  c_S^2  + 2 \rho \beta^{\prime 2} &  2  \rho' \beta^{\prime 2} \sqrt{1-\frac{c_S^2}{\beta^{\prime 2}}} \\
-2 \rho \beta^2 \sqrt{1-\frac{c_S^2}{\alpha^2}} & \rho \beta^2 (2-\frac{c_S^2}{\beta^2}) & -2 \rho' \beta^{\prime 2} \sqrt{1-\frac{c_S^2}{\alpha^{\prime 2}}} & -\rho' \beta^{\prime 2} (2 - \frac{c_S^2}{\beta^{\prime 2}}) \end{vmatrix} = 0. 
\end{equation}
The existence of a root $c_S$ to this equation was first considered by Stoneley \cite{cite:Stoneley_1}, where existence was shown for some special values of the choice of Lam\'e parameters and densities of the half - planes. Further investigations are discussed in \cite{cite:c2}, pp 111 - 113 and the references therein. It has been shown that a root corresponding to the correct Stoneley phase speed can exist when $\frac{\beta}{\beta'} \approx 1$ and that $c_S < \min\{\beta, \beta'\} < \min\{\alpha, \alpha'\}$ always. Hence, the solutions \eqref{eq:s4e3e1} and \eqref{eq:s4e3e2}, if they exist, represent waves traveling harmonically along the interface between the half - planes and decays exponentially into the domain.  

\begin{figure}
	\centering
	\includegraphics[scale = 0.25]{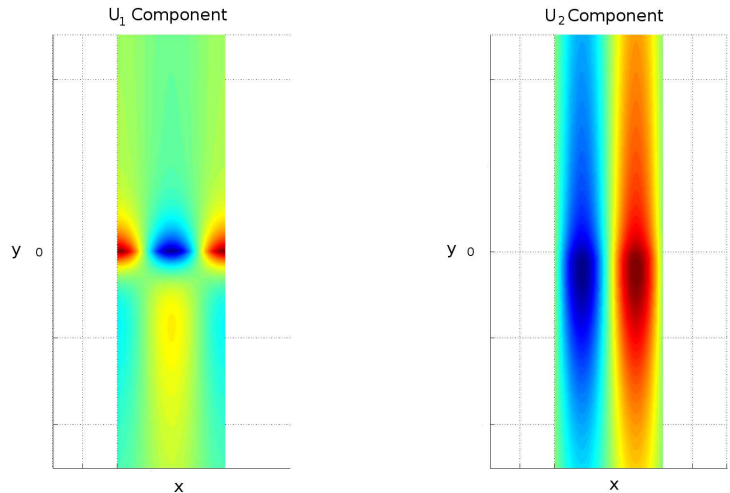}
	\caption{The Stoneley surface wave as a function of $(x,y)$ at $t=0$. The $u_1$ component is show to the left and the $u_2$ component to the right.}
	\label{fig:s4e3f2}
\end{figure}

Numerical simulations involving the Rayleigh surface wave was discussed in \cite{cite:Kreiss_2} and later in the SBP-SAT context in \cite{Virta_2}. There difficulties of the simulation of surface waves was encountered, in particular as the material becomes almost incompressible i.e., $\frac{\lambda}{\mu} \rightarrow \infty$. The purpose of this experiment is to evaluate the performance of the current method when simulating Stoneley waves and to further study the difference in accuracy between schemes using 6th order compatible and fully compatible operators observed in Experiment \ref{exp:e1}. In this study both half - planes are far from incompressible. We take $\lambda = \mu = \rho = 1$ and $\lambda' = 3, \mu' = 2, \rho' = 1.98$. For this case a root  
\[
	c_S =  0.999240140585103
\]
to \eqref{eq:s4e3e3} is found. Figure \ref{fig:s4e3f2} displays vertical and horizontal components of the Stoneley interface wave for this choice of parameters. The computational domain is taken to contain exactly one wavelength, $2\pi$, in the $x$ - direction. Periodic boundary boundary conditions are imposed at $x=0$ and $x=2\pi$ and exact Dirichlet data is imposed at $y = 4 \pi$ and $y = -4\pi$. Initial data for the computations is taken from \eqref{eq:s4e3e1} and \eqref{eq:s4e3e2} at time $t = 0$. We use a grid size of $h = 2\pi/(N-1)$ for $N = 21,41,81,161$ and the max error in the approximate solutions are computed as a function of time for 5 temporal periods. The temporal period is given by
\[
	2 \pi/c_S \approx 6.29.
\]  
\begin{figure}
	\centering
	\includegraphics[scale = 0.25]{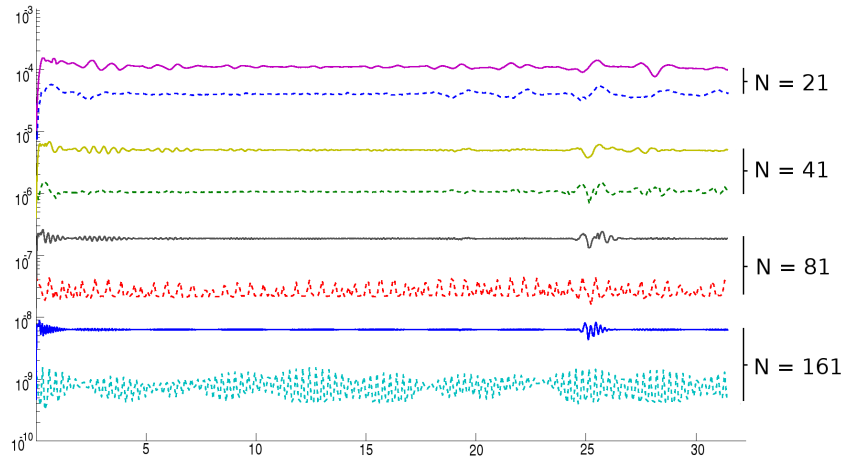}
	\caption{Max errors as functions of time. A dashed line corresponds to the errors obtained with a scheme using fully compatible 6th order operators. A full line corresponds to the errors obtained with a scheme using compatible 6th order operators.}
	\label{fig:s4e3f1}
\end{figure}

The computations are done for each $N$ and each of the two schemes. Figure \ref{fig:s4e3f1} compares the errors obtained with each scheme. A dashed and full line corresponds to the scheme using fully compatible SBP operators and compatible SBP operators, respectively. It becomes evident that the scheme using fully compatible operators converges faster and has a smaller error. For $N = 161$ the error is more than one magnitude smaller.    
\section{Conclusions}
\label{sec:s5}
High order accurate finite difference schemes for the two - dimensional elastic wave equation in a domain consisting of two elastic half - planes in contact have been presented. The key ingredient in the discretization was fully compatible or compatible finite - difference operators approximating first and second derivatives. Stability of the numerical scheme was proven for the discretization using fully compatible operators. Stability for the schemes using compatible operators was only observed in numerical experiments. However, the convergence and accuracy properties was observed to be identical or in some cases better for the case of fully compatible operators. The numerical experiments served to verify convergence, stability and accuracy as well as study the performance of the proposed method on key features arising from a line interface such as mode conversion, refraction arrivals and Stoneley interface waves.

The current method needs several generalizations to be useful in practice. Even though the theory allows for general curvilinear coordinates and isotropic media, the computations were all done i Cartesian coordinates, piecewise constant anisotropic media and two spatial dimensions. A generalization to three spatial dimensions, complicated domains and general media should not pose a theoretical problem, but would require more programming work. An open question in the SBP - SAT context is how to treat grid refinement for equations containing second derivatives in space. A study involving grid refinement for first order systems show good results using SBP preserving interpolation operators \cite{cite:Mattsson_1} but problems has been encountered when applied to higher order equations \cite{cite:GustafssonM_1}. For the method to be significantly more efficient grid refinement is a key feature. Also, an analytical study of the behavior of simulations involving Stoneley interface waves in almost incompressible media would be interesting, as it was seen in \cite{cite:Kreiss_2} and \cite{Virta_2} that the similar Rayleigh suffers increasingly from truncation errors as the media becomes more incompressible.         

\end{document}